\documentclass[aps,pre,showpacs,12pt]{revtex4-1}
\usepackage{graphicx}
\usepackage{xcolor}
\usepackage{bm,amsmath,amssymb}
\newcommand{\bos}{\boldsymbol}

\begin{document}
\title{A contact map method to capture the features of knot conformations }
\author{Neda Abbasi Taklimi$^1$}
\email{neda.abbasi\_taklimi@phd.usz.edu.pl}
\author{Franco Ferrari$^1$}
\email{franco@feynman.fiz.univ.szczecin.pl}
\author{Marcin Rados{\l}aw Pi\c{a}tek$^1$}
\email{marcin.piatek@usz.edu.pl}
\author{Luca Tubiana$^{2.3.4}$}
\email{luca.tubiana@unitn.it}
\affiliation{$^1$CASA* and Institute of Physics, University of Szczecin,
  Szczecin, Poland} 
\affiliation{$^2$ Physics Department, University of Trento, Via
  Sommarive 14, I-38123, Trento, Italy}
\affiliation{$^3$ INFN-TIFPA, Trento Institute for Fundamental Physics and Applications, I-38123 Trento, Italy}
\affiliation{$^4$ Faculty of Physics, University of Vienna, Boltzmanngasse 5, 1090 Vienna, Austria}
\date{\today}

\begin{abstract}
  Inspired by recent advances in the chromosome capture techniques, a method is proposed to study the  structural organization of systems of polymers rings  with topological constraints. 
To this purpose, the system is divided into compartments and
a simple condition
is provided in order to determine if two compartments are in contact or not.
Next, a set of contact matrices $\bar T_{ab}$ is defined that count how many times during a simulation a compartment $a$ was found in contact with a non-contiguous compartment $b$ in conformations with a given energy or temperature.
Similar strategies based on correlation maps have been  applied to the study of knotted polymers in the recent past. The advantage of the present approach is that is coupled with the Wang-Landau algorithm. Once the density of states is computed, it is possible to
generate the contact matrices at any temperature. This gives an immediate overview over
the changes of phases that polymer systems undergo.
The information on the structure of knotted polymers and links stored in the contact matrices  is the result of averaging hundred of billions of conformations
and visualized by means of colormaps.
The obtained color patterns allow to identify the main properties of the structure of the system under investigation at any temperature.
The method is applied to detect the structural rearrangements following the phase transitions of a knotted polymer ring and a circular polycatenane composed by four rings in a solution.
It is shown that the colormaps have a finite number of  patterns that can be clearly associated with the different phases of these systems. The results agree with the available data coming from the plots of the observables and the close inspection of snapshots of the system taken at different steps of the simulations. They also bring new knowledge, for instance predicting the average number of tails appearing in the conformations of the considered polymers at a given temperature.

\end{abstract}
\maketitle
\section{Introduction}\label{introd}
In this work a method is presented to measure the frequency of the interactions between the segments of a system consisting of one or more polymers.
The method is inspired by the recent advances in the techniques that capture the conformations of the chromosomes like the Hi-C technique \cite{Hi-C}.
Each polymer of the system is divided into  compartments containing a number $n_c$ of monomers. A simple criterion to decide if two compartments $a$ and $b$ are getting close is provided: Namely, the distance between their centers of mass should be smaller than the sum of their radii of gyration.
A set of contact matrices $M(E)$ is defined, whose elements $M_{ab}(E)$ count the total number of times in which any pair of compartments $a,b$ of a conformation of energy $E$ has satisfied this criterion during a simulation of the analyzed polymer system. 
In our setup, the polymers are coarse grained and defined on a simple cubic lattice. The random sampling of their conformations is performed in the microcanonical ensemble using the Wang-Landau Monte Carlo algorithm \cite{wl} that allows to compute the density of states $GE(E)$ for any value of the energy $E$.
It turns out that the contact matrices are an useful tool in order to understand the structural organization of a polymer system at the given energy $E$.
By passing from the microcanonical ensemble to the canonical ensemble, it is also possible to study the structural rearrangements following a phase transition.
The method presented in this work is able to capture the main features of states of given energy or temperature after averaging over hundred of billions of sampled conformations. To analyze  such a wealth of data using more traditional tools could slow down considerably the simulations.
Here it is just sufficient to look at the darker or brighter tones of the colormaps generated from the contact matrices. Brighter or darker tones occur
depending on the frequency with which two compartments were found to be close during a given simulation.

The validity of our approach has been checked by applying it to  knotted polymers \cite{edwards,degennes} and circular polycatenanes \cite{LTFFEO2022,liu} in a solution. 
We present here two study-cases: a knotted polymer with the topology of a $4_1$ knot and a circular $[4]$catenane.
The phase transitions of these systems are detected by looking at the plots of their specific heat capacity with respect to the temperature.
The main changes of the structures characterizing the different phases of the considered polymers are determined using the colormaps obtained from the contact matrices computed at temperatures that are lower or higher than the temperature at which a given phase transition is occurring. Conformations stored randomly during the simulations are inspected in order to
verify that the properties predicted from the colormap are frequently occurring as it is expected.

Concluding, we would like to mention some of the previous studies that use other approaches but  are relevant for this work.
The statistical mechanics of open or circular
copolymers has been extensively investigated in the past, see e.~g.
\cite{marko,vilgis,holystvilgis,huber,metzler,binder,kudaibergenov}. Systems
similar to those treated here  have been
considered in \cite{velyaminov,kardar} and,
more recently, in \cite{mella,mella2}. There has been also some interest on circular
diblock copolymers with non-trivial topologies
\cite{orlandinibaiesizontaworkonstiffness,daietal,vlahosetal,najafi,
  kuriatasikorski,benahmed,mella,kumar}. 
Some more general
systems have been considered in relation to specific aspects, like for instance the knotted Hydrophilic-Polar (HP)  models in proteins \cite{dill,wuest,wuest2}, knotted proteins on the lattice~\cite{achille,patricia} and the self-assembly of  nanomaterials of specific topologies controlled by tuning the properties of patchy heteropolymers~\cite{ivan}. 
Previous studies of the statistical
mechanics of knotted  homopolymers and copolymers
with the help of the Wang-Landau algorithm \cite{wl} can be found for example
in \cite{velyaminovhomopolymers,swetnam} (homopolymers) and
 \cite{wang,NATFFMPLT2023} (copolymers).
 There exist also sophysticated alternative methods to investigate the structure of knots.
 For example, Kymoknot \cite{kymoknot}is  a C~code to identify and localize knots based on the Alexander polynomial and Topoly is a Python package to characterize the topology of proteins \cite{topoly}. The HOMFLYPT,
Kauffman, and Jones polynomials are implemented in Knotplot \cite{knotplot}, Topoly, in the Mathematica KnotData
package, and in Python Sage \cite{sage}.
An approach similar in spirit to that discussed here has been recently applied to the study of the dynamics of prime knots~\cite{lappala}. The algorithm that has been used in that work to generate the contact matrices is based on the correlations between the root-mean-squared fluctuations of individual particles of a knot. In our approach not individual particles, but compartments are considered.  The size of the compartments allows to change the resolution under which the knotted polymers or links are investigated.
Another difference is that the present method has been combined with the Wang-Landau algorithm and is specialized to the study of the phase transitions. More in general, it is able to detect the properties of polymer systems in presence of topological constraints at any temperature.

The material presented in this work has been divided as follows.
Section~\ref{method} is an introduction to the used methodology. In particular,
in Subsection~\ref{wala} the main features of the Wang-Landau Monte Carlo algorithm are explained, while Subsection~\ref{hic} is dedicated to the construction of the contact matrix and to the visualization of the data stored in it using colormaps.
The proposed method is then applied in Section~\ref{apps} in order to understand the structural rearrangements following the phase transitions of a single knotted polymer (Subsection~\ref{41}) and a polycatenane composed by four concatenated rings (Subsection~\ref{4cat}).
\section{Methodology}\label{method}
\subsection{The Wang-Landau method used}\label{wala}
The method presented in this work has been tested using
as a model polymer rings in a solution. The
monomers are located on the  sites of a simple cubic lattice and each lattice 
site can be occupied by at most one monomer. Two consecutive monomers
on the loop are linked by one lattice bond,
so that the total length of the knot in lattice
units is equal to $N$.
The polymers considered here are diblock copolymers with $N_A$
monomers of type $A$
and $N_B$ monomers of type $B$. Of course, $N_A+N_B=N$.
The short-range interactions between the monomers are described by following Hamiltonian:
\begin{equation}
  H(X)=\varepsilon(t_{AA}m_{AA}+t_{BB}m_{BB}+t_{AB} m_{AB})
  \label{hamI}
\end{equation}
In Eq.~(\ref{hamI}) $X$ is an arbitrary conformation of the system.
For a given conformation $X$, the quantities
$m_{MM'}$'s, where 
$M,M'=A,B$, count the number of couples composed by a monomer $i$ of type
$M$ and a monomer $j$ of type $M'$ satisfying the following relations:
\begin{equation}
i\ne j\pm1 \quad {\rm and}\quad |\bos R_i-\bos R_j|=1\label{moncond}
\end{equation}
Here $\bos R_1,\ldots,\bos R_N$ denote the locations of the $N$ monomers
and the indices  $i$ and $j$ take all values from $1$ to $N$.
The first condition of Eq.~(\ref{moncond}) is due to the fact that two contiguous monomers along the backbone of the polymer are not interacting.
$\varepsilon$ is an energy scale measuring the cost for two non-contiguous monomers $i$ and $j$ to be found at the minimal allowed distance:
$|\bos R_i-\bos R_j|=1$.
$\varepsilon$ can be positive or negative.
Finally, the $t_{MM'}$ are coefficients that can take
only three values: $0$ and $\pm 1$.
Two monomers $i$ and $j$ of types $M$ and $M'$ respectively are said to form a bond whenever
$t_{MM'}=-1$ and the two conditions (\ref{moncond}) are satisfied.
They $t_{MM'}'$s are used together with
$\varepsilon$ to determine the setup.
For instance, by putting $t_{AA}=t_{BB}=-t_{AB}=1$, we obtain a system of
charged monomers, with the monomers of type $A$ having opposite charge
with respect to monomers of type $B$. In this case, the Hamiltonian
(\ref{hamI})
describes the case discussed in \cite{NATFFMPLT2023}. of Coulomb interactions screened by the presence of ions in the
solution.
Alternatively, choosing $t_{AA}=1$, $t_{BB}=t_{AB}=0$ and $N_A=N,
N_B=0$, the knotted polymer becomes an homopolymer in a good solvent
for $\varepsilon>0$ and in a bad solvent for $\varepsilon<0$.
The used setups are summarized in Table~\ref{table1}.
In that table, the case diblock copolymers~I refers to the charged
monomers mentioned before. In diblock copolymers~II the solvent is
good for the monomers of type $A$ and bad for
those of type $B$.
\begin{table}
 \begin{tabular}{|c|c|c|c|c|c|c|}\hline
setup  &$\varepsilon$&$t_{AA}$&$t_{BB}$&$t_{AB}$&$N_A$&$N_B$\\\hline
homopolymers  &$>0$&$1$&$0$&$0$&$N_A=N$&$N_B=0$\\
in good solvent  &&&&&&\\\hline
homopolymers  &$<0$&$1$&$0$&$0$&$N_A=N$&$N_B=0$\\
in a bad solvent  &&&&&&\\\hline
diblock  &$>0$&$1$&$1$&$-1$&$N_A>0$&$N_B=N-N_A$\\
copolymers I  &&&&&&\\\hline
diblock  &$>0$&$1$&$-1$&$0$&$N_A>0$&$N_B=N-N_A$\\
copolymers II  &&&&&&\\\hline
   \end{tabular}
  \caption{Parameters describing the setups discussed in this work. }\label{table1}
  \end{table}
For convenience,
thermodynamic units will be chosen in which the Boltzmann constant
is equal to one. In these units the temperature $\theta$ is related to
the usual temperature $T$ by the relation: $k_BT=\theta$.
We will also introduce the rescaled temperature
$ \bar T=\frac \theta\varepsilon$. Clearly, the ratio $\frac
{H(X)}{k_BT}=\frac{\bar H(X)}{\bar T}$, where $\bar
H(X)=H(X)/\varepsilon$:
\begin{equation}
    \bar H(X)=t_{AA}m_{AA}+t_{BB}m_{BB}+t_{AB} m_{AB}\label{redham}
  \end{equation}
More details are given in Ref.~\cite{NATFFMPLT2023}.

The simulations are performed using the Wang-Landau Monte Carlo
algorithm \cite{wl}.
The initial knot conformations are obtained by elongating
the existing conformations of minimal length knots \cite{rechnitzer,sharein} until the desired final length is attained.
The details on the sampling and the treatment of the
topological constraints can be found in
Refs.~\cite{yzff} and \cite{yzff2013}.
The random transformations that are necessary for sampling the different knot conformations are the pivot moves of Ref.~\cite{madrasetal}.
In order to preserve the topological state of the system,  the pivot algorithm and excluded area (PAEA) method of Ref.~\cite{yzff} is applied. 

The partition function of the polymer knot is 
given by:
\begin{equation}
Z(\bar T)=\sum_{ E}e^{-E/{\bar T}}g(E)
\end{equation}
where $g(E)$ denotes the density of
states:
\begin{equation}
g(E)=\sum_X\delta(H(X)-{E})
\end{equation}
$g(E)$ is
the quantity to be evaluated via Monte Carlo methods.
The
expectation values of any observable $\cal O$ may be computed
using the formula:
\begin{equation}
  \langle{\cal O}\rangle(\bar T)=\frac 1{Z(\bar T)}\sum_{ E}e^{-E/\bar   T}g(E){\cal O}_{ E}\label{expval}
\end{equation}
Here ${\cal O}_E$ denotes the average of $\cal O$ over all
sampled states with rescaled energy $E$.
\subsection{The contact matrix}\label{hic}
To construct the contact matrix, the polymer under investigation is divided into
"compartments" with $n_c$ monomers each. Supposing that the polymer has a total of $N$ monomers, the number of compartments is $N/n_c$.

\begin{figure}[h]
  \begin{center}
\includegraphics[width=1.0\textwidth]{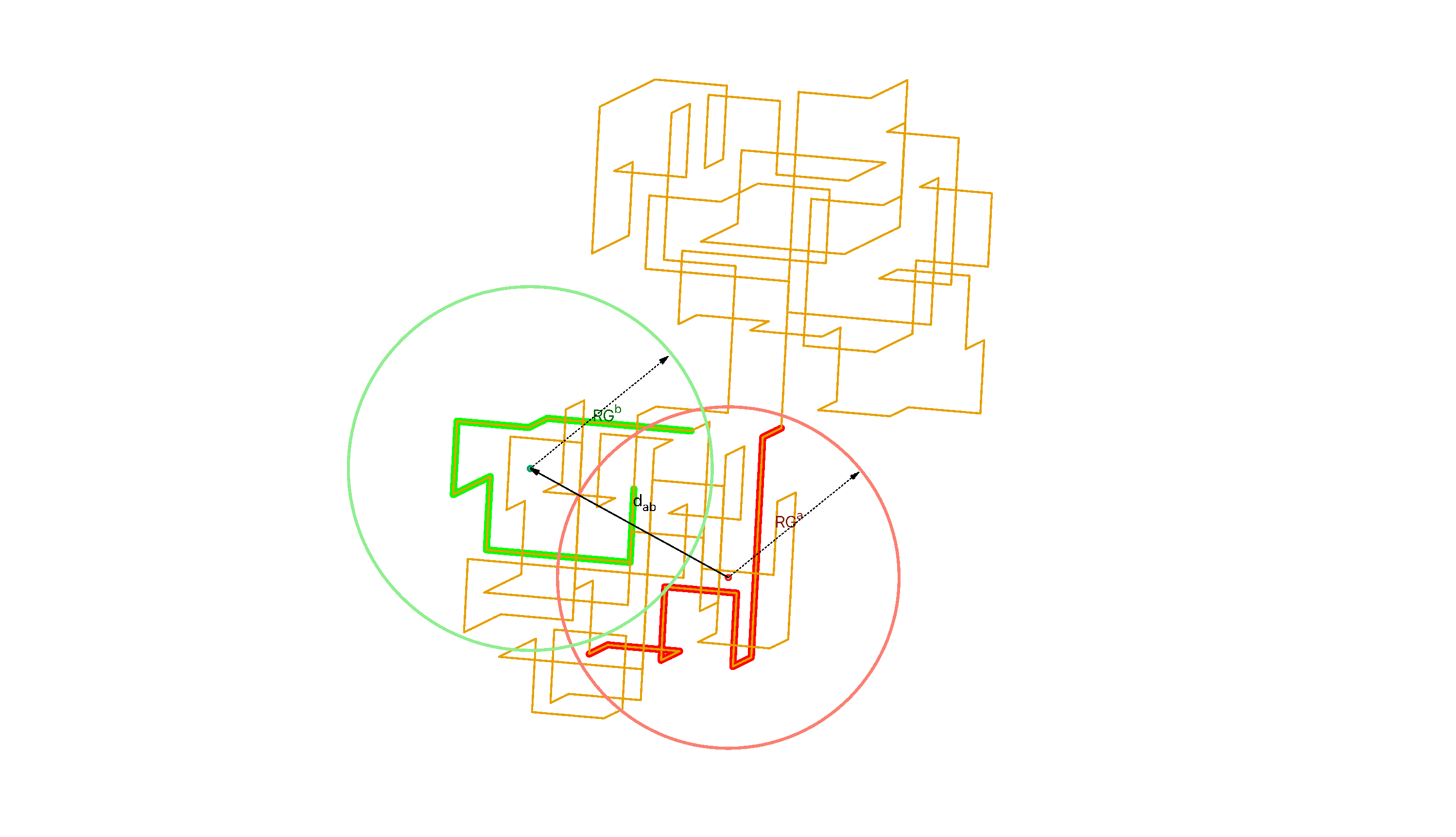}
  \end{center}
  \caption
{In this picture one conformation is shown of a system obtained by joining
    together the ends of two open knots using two chains. The
    resulting topology is that of a composite knot of total length
    $N=192$ consisting of a
    knot $3_1$ (in the left upper part of the picture) with $N_{3_1}=24$ and
    a knot $6_1$ (in the right lower
    part of the picture) of length
  $N_{6_1}=80$. The ends of both knots are crosslinked in such a way that, if we take
    $n_c=6$,    the knots $6_1$ and $3_1$ are strictly confined within
  the compartments $1-14$ and $20-25$ respectively.
The knot $3_1$ is of minimal length and, for this reason, it cannot fluctuate.
In the picture two compartments $a$ and $b$, respectively in red and
green colors, have been singled out. According to Eq.~(\ref{contactcondition}),
two compartments $a$ and $b$ are in contact 
    provided that  the distance $d_{ab}$ between their
    centers of mass is less than the sum of their gyration radii $R_G^{(a)}+R_G^{(b)}$.
    Essentially, this means that the  volumes occupied by the two
    compartments  are overlapping. 
  }
  \label{condmeaning}
  \end{figure}

Let's label these compartments with the first letters of the Latin alphabet
$a,b,c,\ldots=1,\ldots,N/n_c$.
During the sampling, every time a compartment $a$ gets "in contact" with another
compartment $b$, the element $ M_{ab}$ of the contact matrix is updated
as follows:
\begin{equation}
M_{ab}=M_{ab}+\mu\label{procedure}
\end{equation}
where $\mu$ is some small real number, for instance $\mu=0.00000001$.
Of course:
\begin{equation}
M_{ab}=M_{ba}\label{symmetry}
\end{equation}
The condition for which two compartments are considered in contact will be provided  in the following. First, the
gyration radii $R_G^{(a)}$ and $R_G^{(b)}$ of the two compartments is
computed.
The two compartments $a$ and $b$ are said to be in contact if the distance
$d_{ab}$ between their centers of mass satisfies the relation:
\begin{equation}
  d_{ab}\le R_G^{(a)}+R_G^{(b)}\label{contactcondition}
\end{equation}
The meaning of Eq.~(\ref{contactcondition}) is illustrated by
Fig.~\ref{condmeaning}.
The above procedure produces the contact matrix $ M_{ab}$
that counts
 how many  times during a simulation a conformation is found such that
 two compartments $a$ and $b$ are in contact for $a,b=1,\ldots,N/{n_c}$.
 The data stored in $M_{ab}$ may be visualized using a colormap.
 For instance, in the colormap
 darker colors can be assigned to pairs of compartments that have resulted to be more frequently distant from each other and
 lighter colors to pairs that were closer. This  convention will be used 
 in all the colormaps presented here.
 
At least in principle, these data 
 deliver an information about the shape of knotted polymers.
However, in a typical simulation hundred of billions of random conformations are generated. These conformations have very different shapes, from extremely compact to extremely expanded and any case inbetween.
It is difficult to capture the average features of such a wealth of
conformations. To obtain meaningful results, a further refinement is
necessary consisting in the introduction of a new contact matrix $M_{ab}(E)$.  $M_{ab}(E)$ is defined exactly like the full matrix $M_{ab}$, but with the restriction that only conformations with a fixed energy $E$ are considered.
Even with this restriction, the variety of shapes is still enormous. Yet,
the performed simulations show that the matrix $M_{ab}(E)$  is able to capture
the common features characterizing the conformations of given energy despite the
fact that the shapes of these conformations strongly differ from each other.
Indeed, we have observed that after a sufficient number of samples $K$ has
been generated, the colormap stabilizes and there are no more
significant changes of the patterns formed by the darker and lighter areas of the map.
Of course, during the sampling the elements $M_{ab}(E)$ are steadily
growing because of the addition of the small quantity $\mu$ in Eq.~(\ref{procedure}),
However,
denoting by
$M_{ab}^{(K)}(E)$ and $M_{ab}^{\lambda K}(E)$ the  contact
matrices computed after taking into account $K$  and $\lambda K$
conformations respectively, we have that:
\begin{equation}
\lim_{K\to \infty}M_{ab}^{(\lambda K)}(E)= \lambda M_{ab}^{(K)}(E)\label{scalingcond}
\end{equation}
where $K>>1$  and $\lambda >1$ is a scaling factor independent of $a$ and $b$.
In other words, when large amounts of conformations have been explored,
the ratio 
\begin{equation}
{\bar M}_{ab}(E)=  M_{ab}^{(\lambda K)}(E)/M_{11}^{(\lambda K)}(E)\label{normmhic}
\end{equation}
converges to a given value which does not change  if further samples
are  considered.
While convergence require $K>>1$, this is not a problem because, during a
simulation, a very high number of conformations is explored, usually
of the order $10^{11}$ or higher.
The conditions (\ref{scalingcond}) and (\ref{normmhic}) have
been verified in all performed simulations.

\begin{figure}[h]
  \begin{center}
    \includegraphics[width=0.30\textwidth]{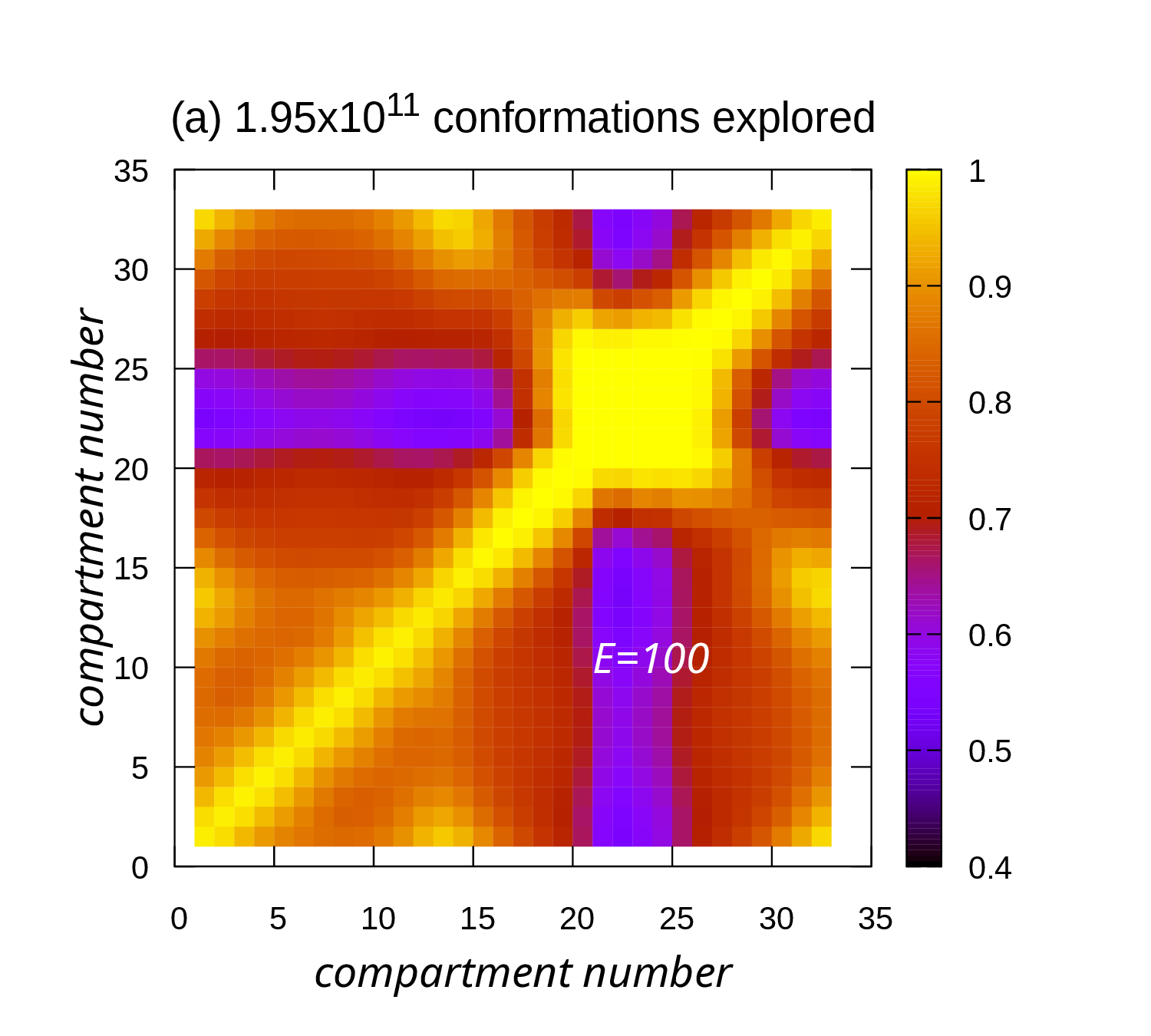}
    \includegraphics[width=0.30\textwidth]{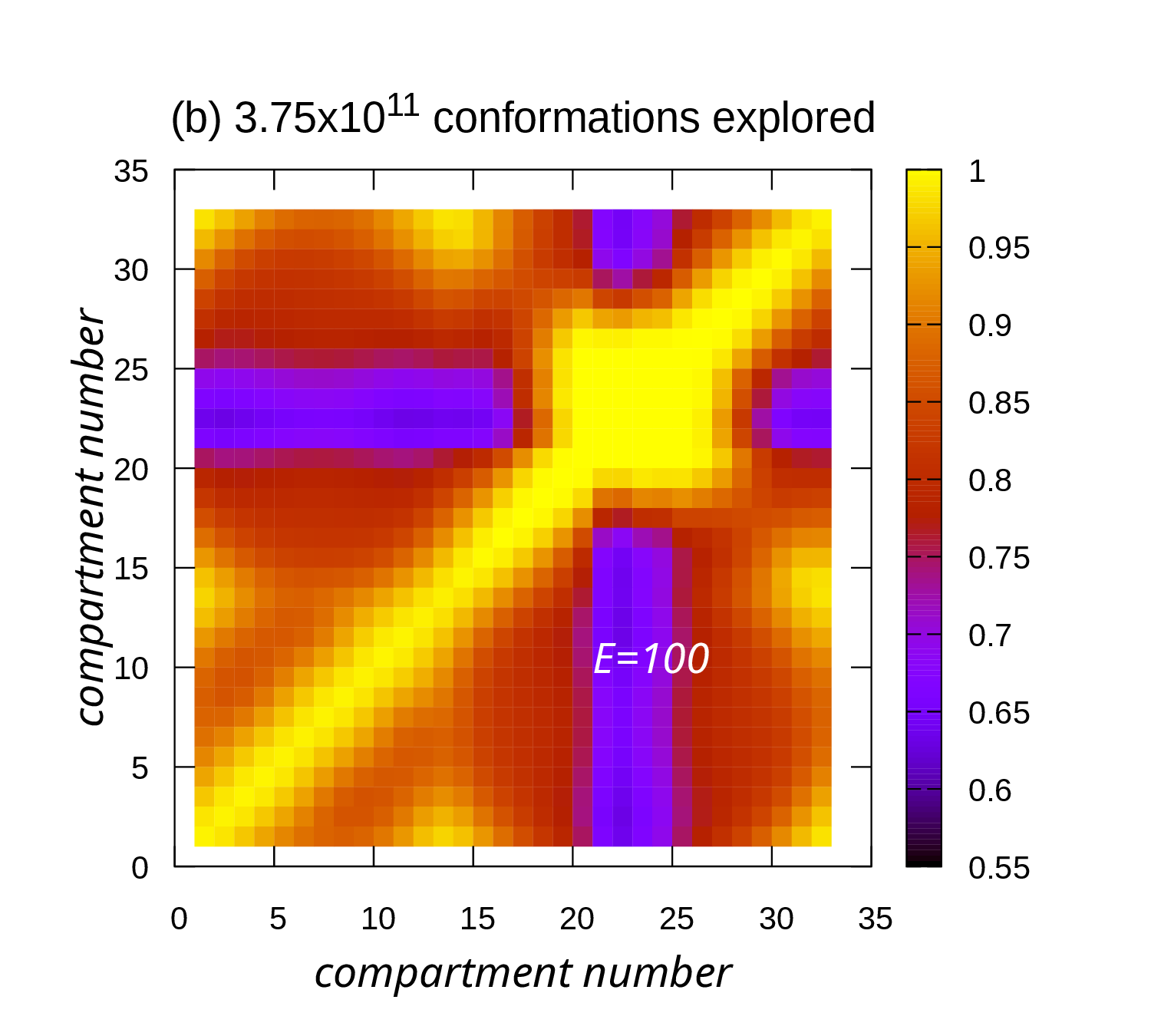}
    \includegraphics[width=0.30\textwidth]{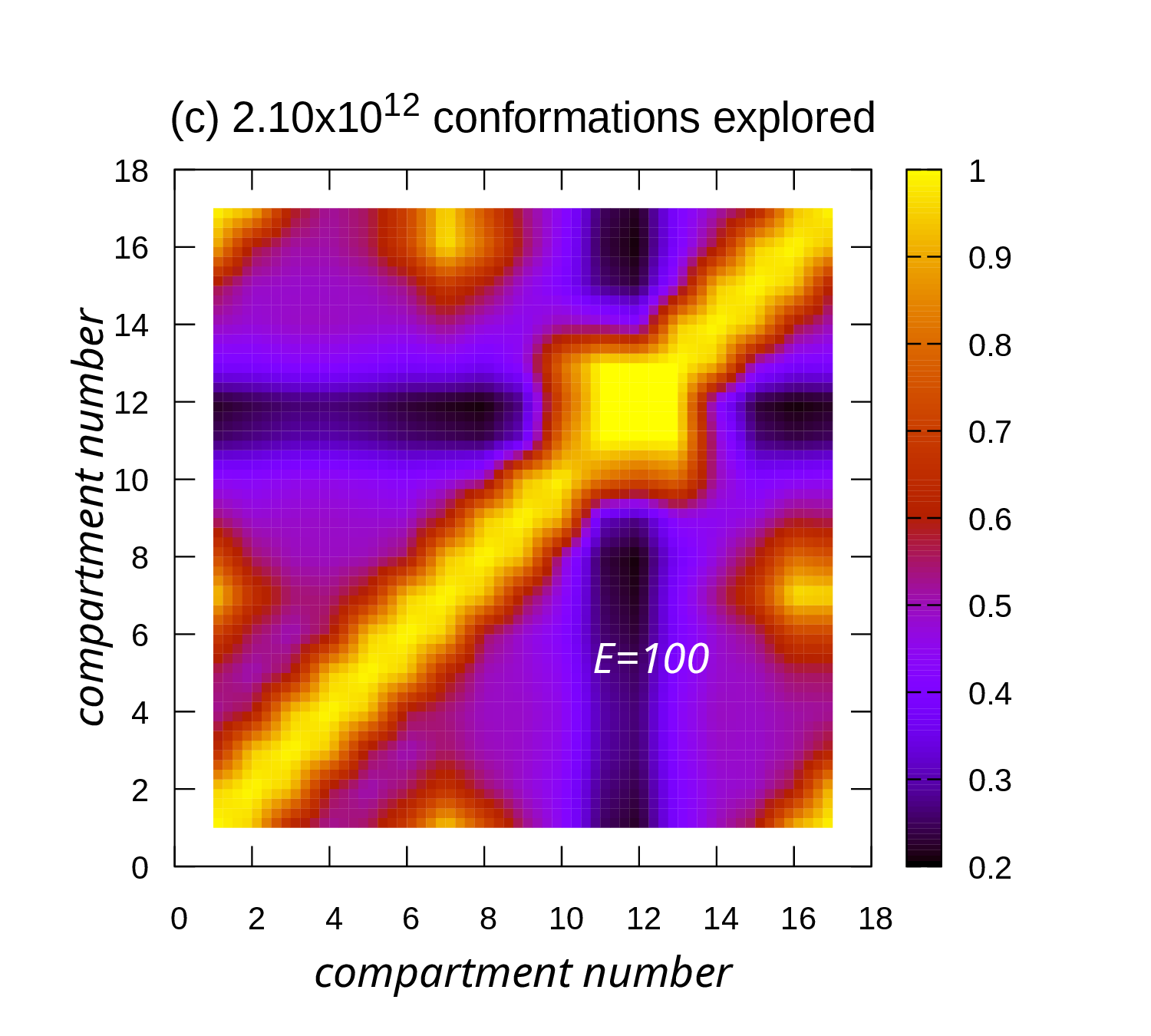}
  \caption{This picture shows two colormaps of the system in
    Fig.~\ref{condmeaning} taken at different instants of the
    simulation, see panels (a) and (b). In panel (c) the third
    colormap has been generated using a lower resolution ($n_c=12$
    instead of $n_c=6$) but increasing the number of interpolated points.
    The fixed energy of the system is $E=100$.
  }\label{colormap} 
  \end{center}
\end{figure}

We would also like to stress that the information stored in the matrices
${\bar M}_{ab}(E)$ is robust with respect to changes in
the choice of $n_c$, i.~e. the parameter that determines the
resolution under which a knotted 
polymer is observed. More precisely, supposing
that $n_c'>n_c$, a colormap generated from a matrix ${\bar
  M}_{ab}(E)$ computed using the lower resolution $n_c'$, but interpolated 
over a larger number of points, presents the same patterns of bright and dark areas as
the colormap obtained  with the higher resolution $n_c$, see Fig.~\ref{colormap}.
 Of course, in the limiting case $n_c=N$, in which the compartment
 coincides with the whole system, the resolution will be too low to
 get some useful information about the polymer conformations from
 the matrix $\bar M_{ab}(E)$.  On the other hand, too small compartments,
 let's say with $n_c<5$, have a too small radius, so that Eq.~(\ref{contactcondition})
can hardly be applied.

As an example to illustrate the above settings, in Fig.~\ref{colormap}  a few
colormaps are shown corresponding to the knotted homopolymer
of Fig. ~\ref{condmeaning}.
The topology of the system is that of a composite knot  $3_1\# 6_1$. The total number of monomers is $N=192$.
The resolution has been set by putting $n_c=6$, so that there are $32$ compartments.
To take into account the fact that
the polymer forms a closed curve in space, a fictitious $33-$compartment has been added whose entries in the contact matrix $\bar T_{ab}(E)$  coincide with those of the first compartment. 
Accordingly, the colormaps are divided into
$33\times 33=1089$ small sectors corresponding to the $1089$ elements of $\bar T_{ab}(E)$. 
Moreover, the small sectors on the opposite edges of the colormaps are identified as in the rectangle representing a torus on a two dimensional plane.
Finally, in the reduced Hamiltonian $\bar H(X)$ of Eq.~(\ref{redham}) describing the energy of the system the parameters $t_{MM'}$ have been chosen according to the setup homopolymer in a good
solvent, see Table~\ref{table1}. More information can be found in the captions of Fig.~\ref{condmeaning}.
We would like to stress the presence of cross-links constraining the knots $3_1$ and $6_1$ to be confined
within the compartments $20-25$ and $1-14$ respectively.

The data used in panels~(a) and (b) come from the matrices
$\bar M_{ab}^{(K)}(E),\bar M_{ab}^{(K')}(E)$  taken after
considering $K=1.95\cdot 10^{11}$  and $K'=3.75\cdot 10^{11}$
conformations respectively.
The energy $E=100$ is the average energy of the system at the temperature $\bar T\sim 1.19$ .
In panel (c) the resolution has been reduced to $n_c=12$, but a higher
interpolation level has been applied.
As it is possible to see, the color
patterns in  panel (a) and (b) are practically the same and they are in a
good agreement with that of panel (c).
Let us notice that the range of values corresponding to the colors in
panel~(a) (from 0.4 to 1) is different from that of 
panel~(b), which goes from 0.55 up to 1.
However, the elements of the matrix $\bar M_{ab}^K(E)$ in
panel~(a) that are within the interval $0.4 - 0.55$
consist of less than 7\% of the total of $32^2=1024$
elements. This is why their presence does not introduce substantial
changes in the colormap of panel ~a) with respect to that of 
 panel~(b). Small discrepancies like this may occur even at very high
values of $K$ because, while the system is ergodic,
certain conformations need a  considerable amount of time before being generated
started from a given seed. 

In the next Section we will use the following notation to identify an arbitrary
sector on a colormap: $x(l:m),y(n:p)$. For instance, the sector $x(21:25),y(1:16)$
coincides with the strip in panel (a) of Fig.~\ref{colormap} in
which the purple color is dominant. This means that the compartments
$21-25$ have a low chance to get in contact with the compartments $1-16$.

\section{Application of the technique: capturing the structure of
  knotted polymers}\label{apps}
The contact matrix method presented in the previous Section is a helpful tool in
 understanding the structure of knotted polymers and the structural reorganizations that they
undergo during  phase transitions.
It is particularly useful  within the Wang-Landau method,
because in that case the sampling is performed in the microcanonical
ensemble, where the temperature $T$ is not known.
This makes
it difficult to take snapshots of the conformations that could show
the structure of knotted polymers
below and above the transition temperature.
Once the matrices $\bar M_{ab}(E)$  defined in the previous Section are computed for all values of the energy $E$,
the matrices $\bar M_{ab}(\bar T)$ in the temperature domain
can be recovered  using Eq.~(\ref{expval}):
\begin{equation}
{M}_{ab}(\bar T)=\frac 1{Z(\bar T)}\sum_{E}e^{-E/\bar   T}g(E)M_{ab}(E)
\end{equation}
$\bar M_{ab}(\bar T)$ counts how
frequently two compartments $a$ and $b$ have been found in contact
at a given temperature $\bar T$ during a simulation.
\subsection{Case of a knotted polymer with topology $4_1$ and $N=200$}\label{41}
\begin{figure}[h]
  \begin{center}
\includegraphics[width=0.70\textwidth]{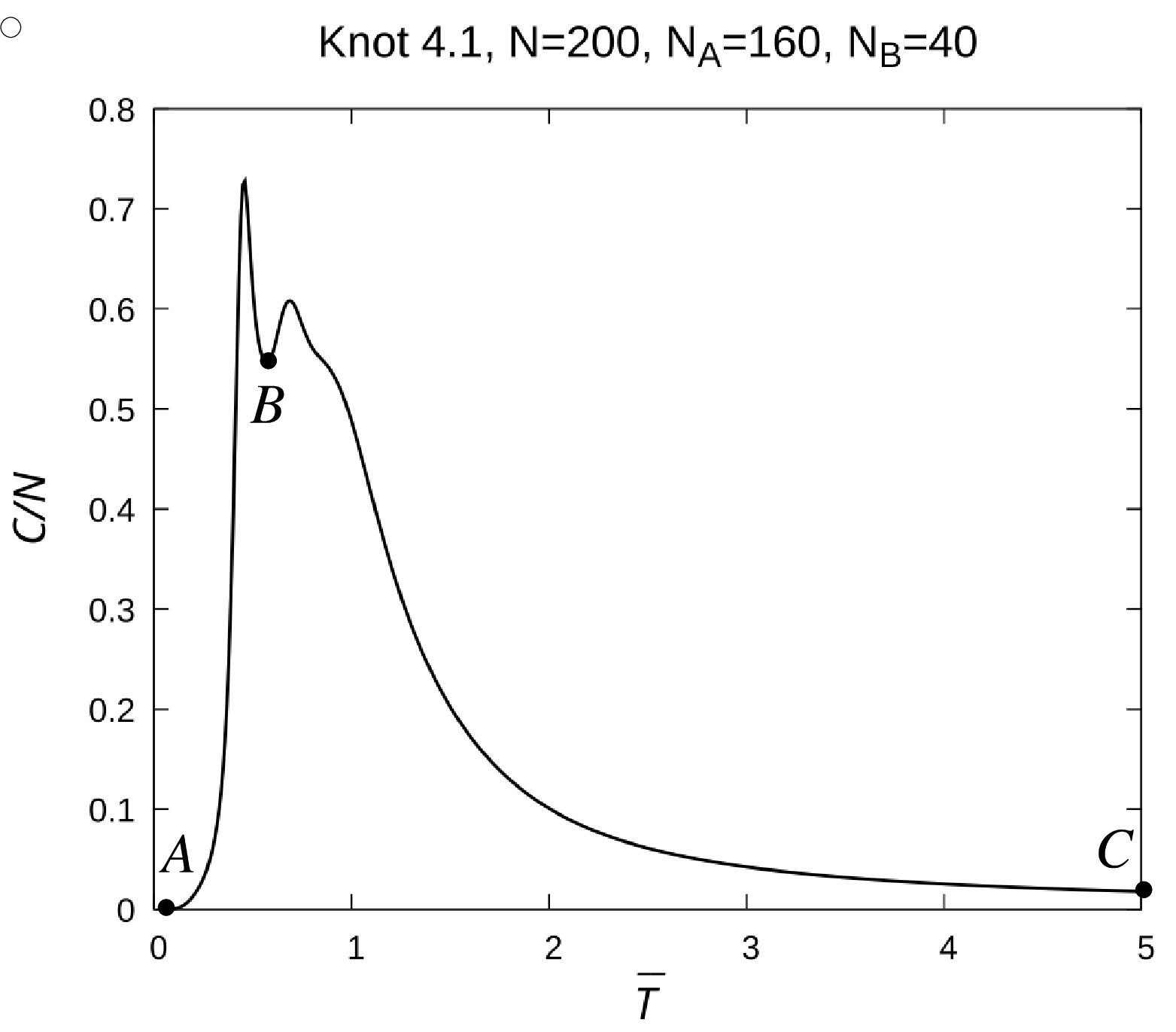}
  \end{center}
  \caption
{In this figure the plot of the specific heat capacity $C/N$ of a diblock
  copolymer ring with the topology of a figure-eight knot $4_1$ is
  presented. $C/N$ exhibits two peaks, corresponding to the two
  transitions from a $AB-$mixed phase $\bar M$ to an intermediate
  phase $\bar I $ and from $\bar I$ to an unmixed phase $\bar U$. In
  the figure the points $A,B$ and $C$ correspond to the respective temperatures
  $T_A=0.05, T_B=0.56$ and $T_C=5.00$ in which  the phases $\bar
  M,\bar I$ and $\bar U$ dominate.
  }
  \label{N200-phase-transition}
\end{figure}

\begin{figure}[h]
  \begin{center}
\includegraphics[width=0.30\textwidth]{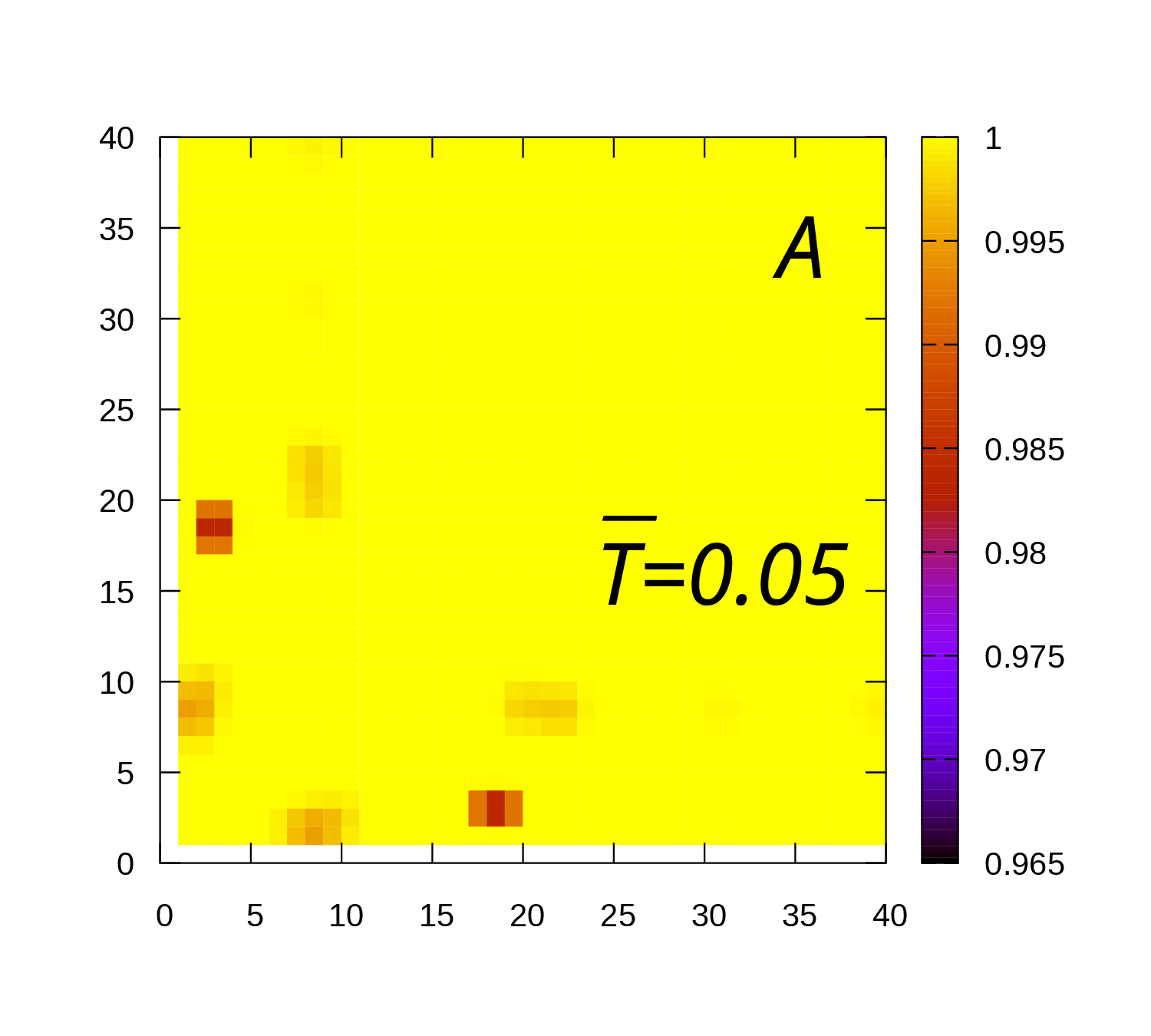}
\includegraphics[width=0.30\textwidth]{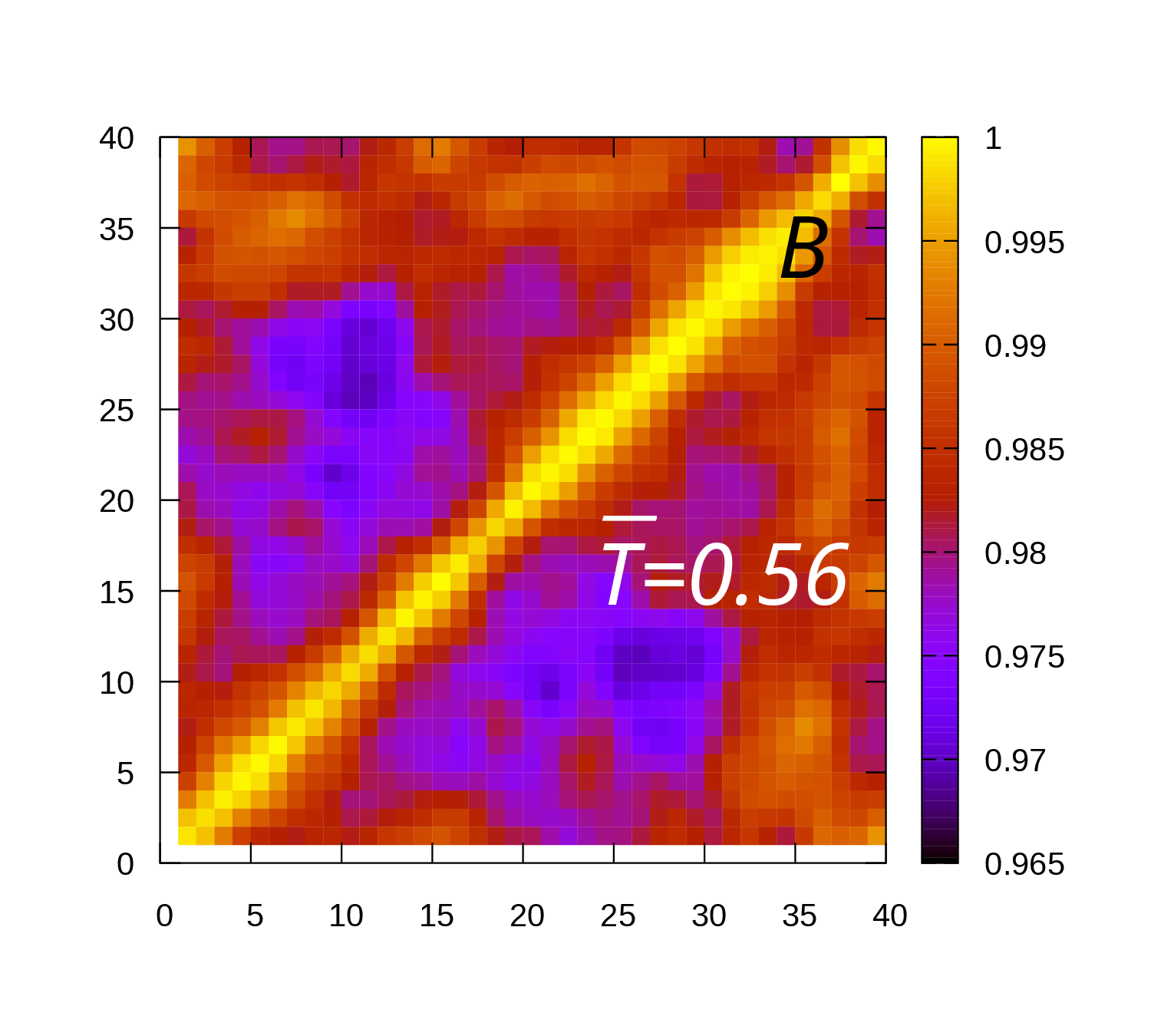}
\includegraphics[width=0.30\textwidth]{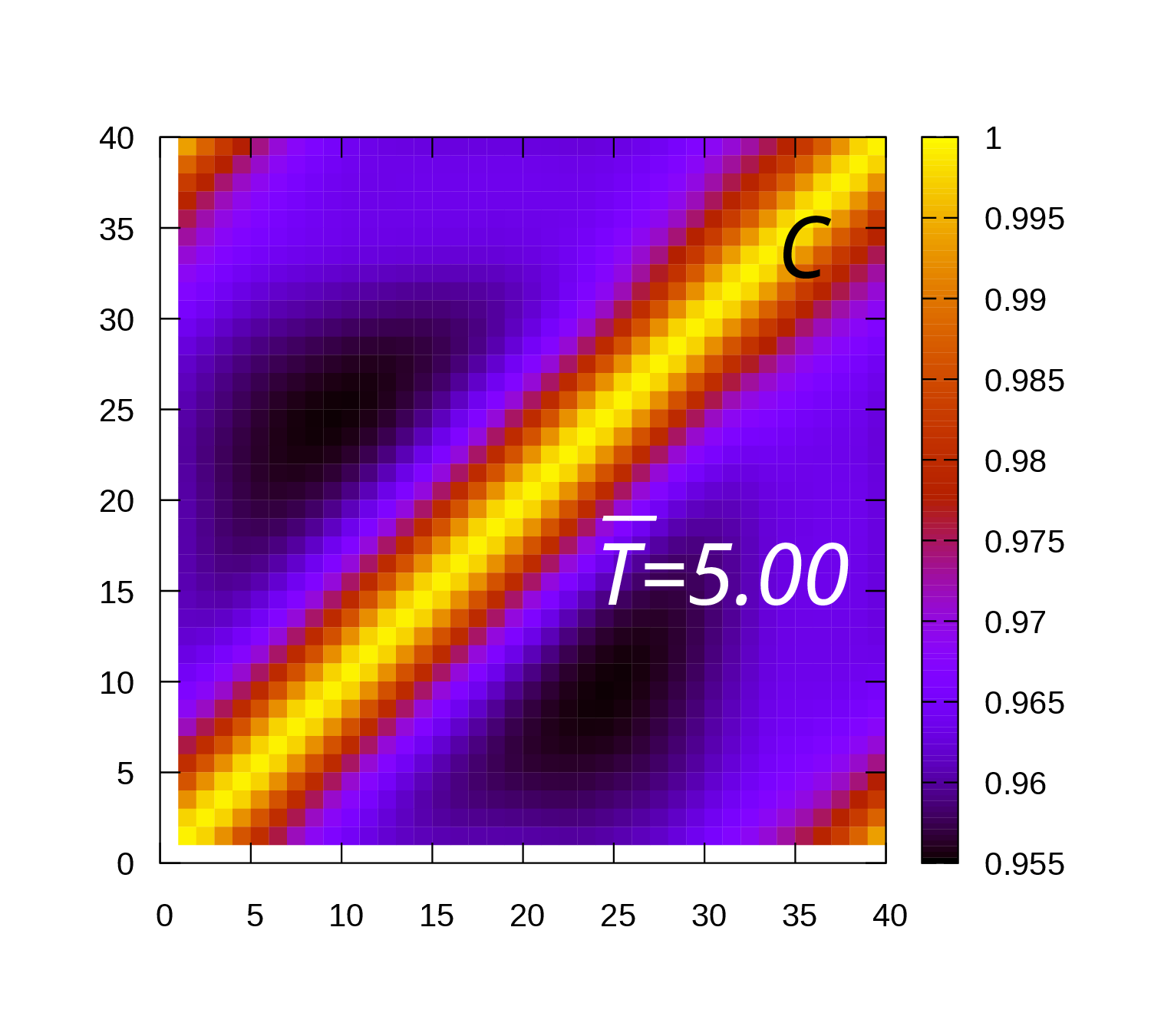}
  \end{center}
  \caption
{The  colormaps in this figure correspond to the three distinct structural
  organizations that
  the
  conformations of a diblock knotted ring with $N=200$ monomers and
  the topology of the figure-eight knot $4_1$ take at different temperatures. The monomer composition of
  the system is $N_A=160$ and $N_B=40$, the setup is that of diblock copolymers~I, see Table~\ref{table1}
  and the caption of Fig.~\ref{N200-phase-transition-sample-conformations}.
  The shown colormaps $A,B$ and $C$ have been computed at the temperatures $\bar T_A=0.05, \bar T_B=0.56$ and $\bar T_C=5.00$ respectively.
  }
  \label{N200-phase-transition-colormaps}
\end{figure}

To show how
an information about the structural organization of knotted polymers
can be retrieved, we
consider as a first example a ring with $N=200$
monomers and the topology of a $4_1$ knot. Let the setup be that of diblock
copolymers~I of Table~\ref{table1} with $N_A=160$ and $N_B=40$.
The number of monomers in a compartment is $n_c=5$, so that each
conformation is divided into $40$ compartments.
The first $32$ compartments contain monomers of type $A$, while the
last eight contain monomers of type $B$.
This system, characterized by an excess of $A$ monomers, is known to exhibit
three distinct phases as explained in~\cite{NATFFMPLT2023}: 
a mixed phase ($\bar M-$phase) at the lowest temperatures, an intermediated phase
($\bar I-$phase) and an unmixed phase ($\bar U-$phase) at high temperatures.
The mixing is between the $A$ and $B$ monomers.
The plot of the specific heat capacity $C/N$ of this system shows two
peaks, see
Fig.~\ref{N200-phase-transition}. They correspond
to the transitions $\bar M\longrightarrow \bar I$
and $\bar I\longrightarrow \bar U$.
In the figure the points $A,B$ and $C$ correspond to the temperatures
$\bar T_A=0.05, \bar T_B=0.56$ and $\bar T_C=5.00$ respectively.
At $\bar T_A$ the $\bar M-$phase dominates, while at $\bar
T_B$ and $\bar T_C$ the systems is in the $\bar I-$ and $\bar U-$phases respectively.
In Fig.~\ref{N200-phase-transition-colormaps} the colormaps $A,B,C$
obtained from the matrices $\bar M_{ab}(\bar T_A),\ldots,\bar
M_{ab}(\bar T_C)$ are shown.
As it is possible to see, in all of them the central
diagonal is yellow, which is the color corresponding to
 the highest possible value in all colorbars. 
This is expected, because it is very likely that two contiguous
compartments are getting in contact.
Another common feature is that the values in the colorbars are ranging within
small intervals, the largest of them being [0.955,1.000] in the right panel of Fig.\ref{N200-phase-transition-colormaps}. This is explained by the fact
that the colormaps are the result of the averaging over
hundred of billions of conformations. As a consequence,
no matter how two compartments $a$ and $b$ are
located along the polymer backbone, it is well possible that
there is a statistically relevant set of conformations in which $a$ and $b$ are close.
For this reason, the differences between the elements $\bar M_{ab}(E)$
can be small. However, as we will see these differences produce colormaps that are able to distinguish the various phases of knotted polymers .
The range of the colorbar is of course sensitive to the presence of constraints.  For instance,
the interval of values
in the colorbars of
Fig.~\ref{colormap} is large because of the crosslinks that limit the movements of the compartments of
the composite knot $3_1\#6_1$ discussed in Section~\ref{method}. An
extended range is likely to occur also when the
set of conformations belonging to a given phase is small.
In this case, the limited number of available conformations could be characterized by a restricted number of  shapes preventing some of the compartments to get close to each other.

\begin{figure}[h]
  \begin{center}
\includegraphics[width=0.30\textwidth]{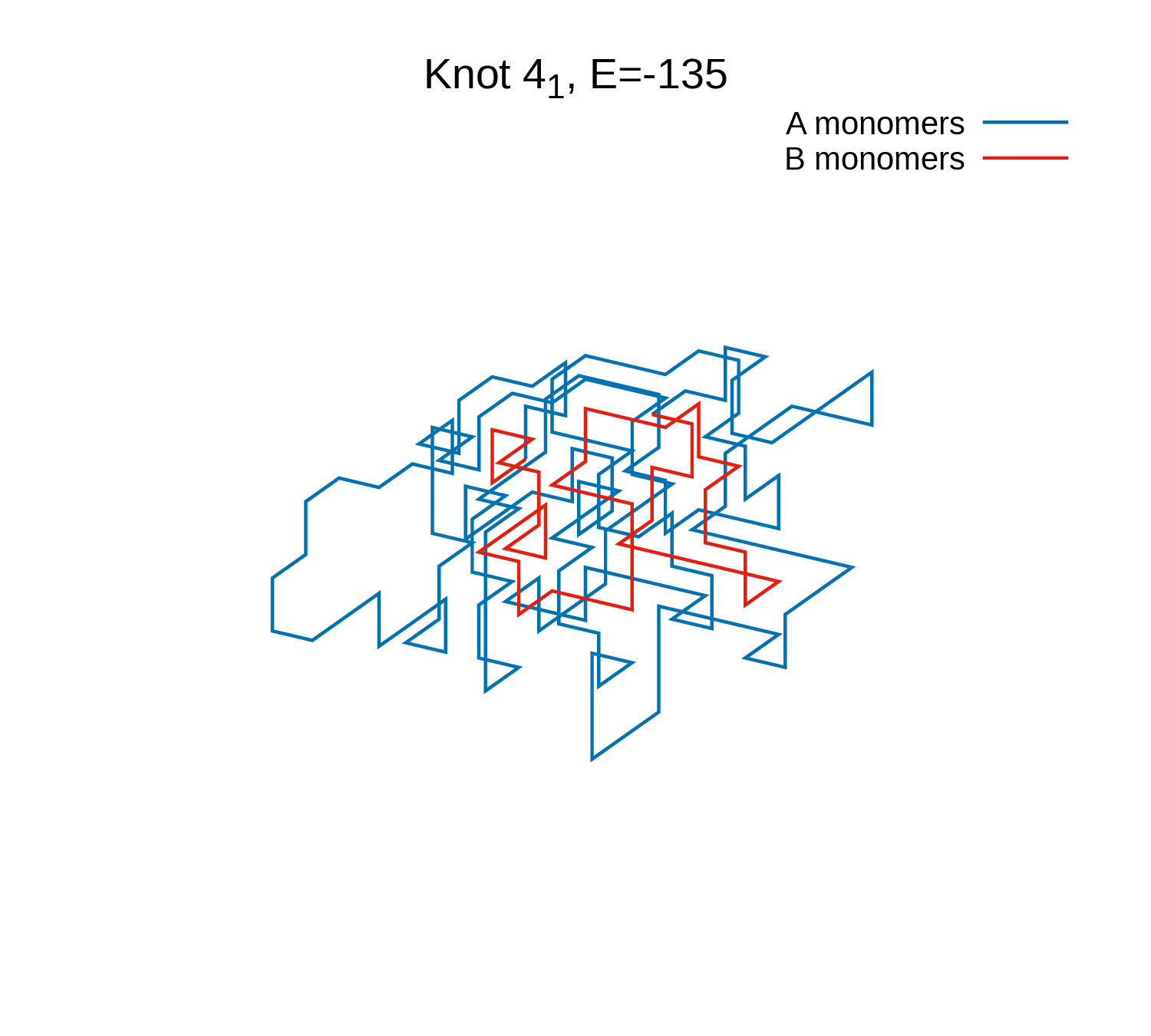}
\includegraphics[width=0.30\textwidth]{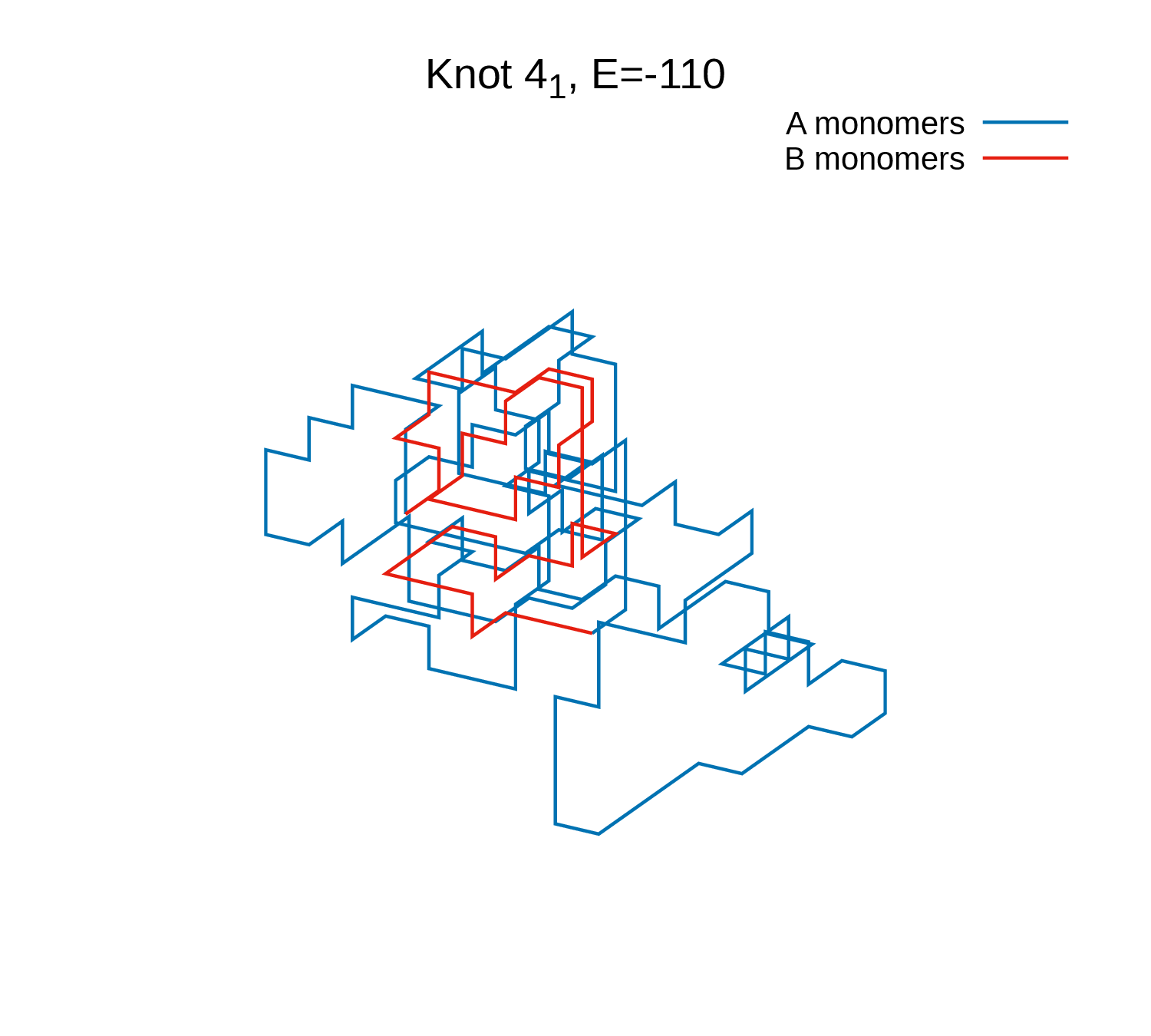}
\includegraphics[width=0.30\textwidth]{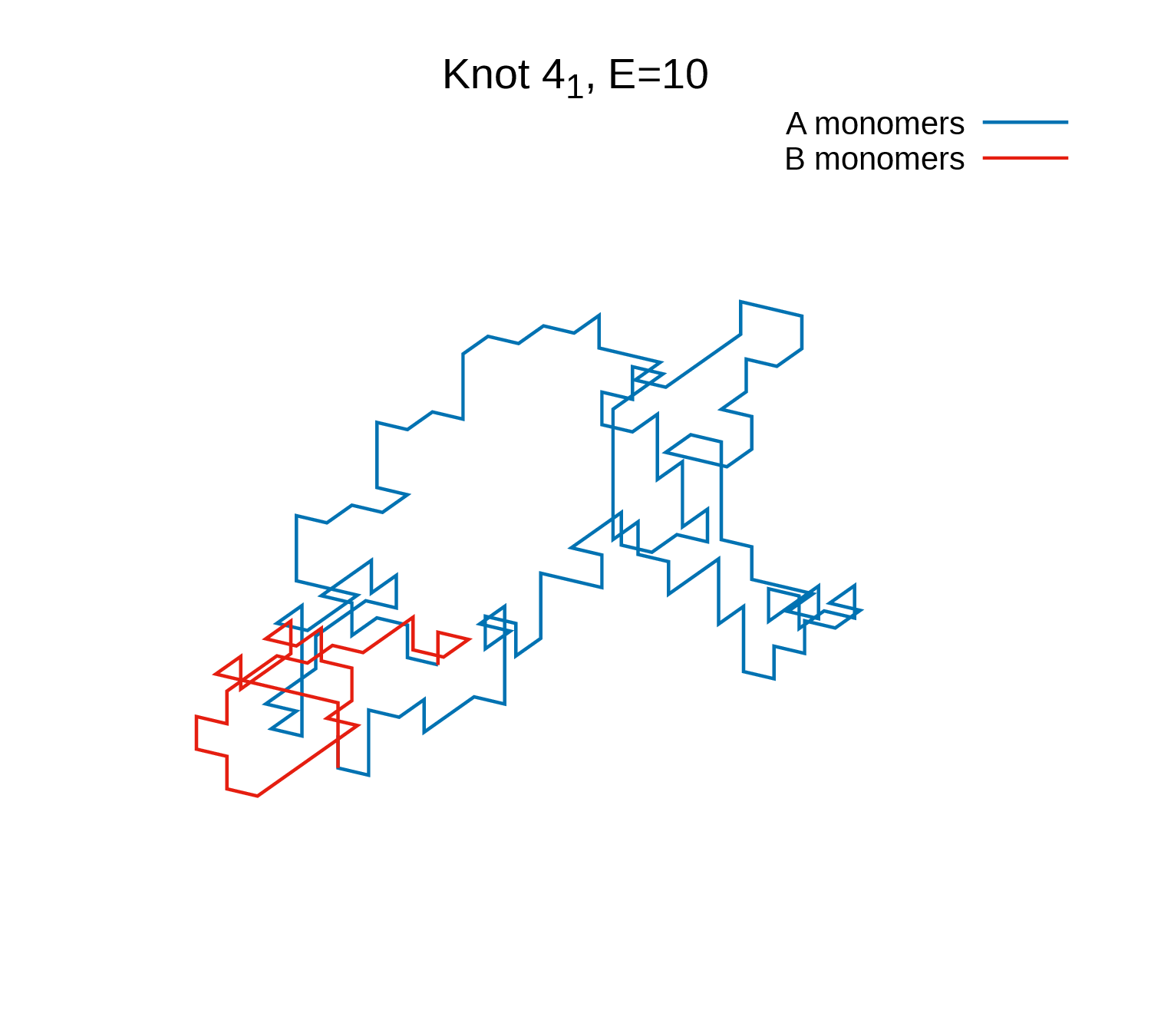}
  \end{center}
  \caption
{Sample conformations with topology $4_1$ that
  are typical of the $\bar M-$phase (left panel),
  $\bar I-$phase (center panel) and the $\bar U-$phase (right panel).
  All conformations have $N=200$ monomers, of which $N_A=160$  are $A$
  monomers and $N_B=40$ are $B$ monomers. The segment with the $A$
  monomers is shown in blue color and the segment with the $B$
  monomers in red color. The conformation in the left panel has
  energy $E=-135$. This is the average  energy of the
  system at the temperature $\bar
  T=\bar T_A=0.05$. Analogously,  the values $E=-110$ and $E=10$ of
  the remaining two conformations are the average energies of the
  knotted ring at $\bar T_B=0.56$ and $\bar T_C=5.00$ respectively.
  }
  \label{N200-phase-transition-sample-conformations}
\end{figure}

Coming back to Fig.~\ref{N200-phase-transition-colormaps},
we note that the colormaps $A,\ldots,C$ 
exhibit three quite distinct patterns.
These patterns
correspond to the 
structural organizations of the system in the three available phases. In fact, they
remain stable if the temperature is increased or decreased
and change only when one of the two peaks of the specific heat capacity is crossed.
Similar colormaps as that in the left panel of Fig.~\ref{N200-phase-transition-colormaps}
are observed at the lowest temperatures in which $0<\bar T\le 0.30$.  The figure shows the colormap in the particular case $\bar T_A=0.05$.
The distribution of darker and brighter colros is compatible with the
  conformations of the $\bar M-$phase
  described  in 
  Ref.~\cite{NATFFMPLT2023}. An example of such conformations,
  corresponding to the lowest possible energy $E=-135$, is provided
  in Fig.~\ref{N200-phase-transition-sample-conformations},
  left panel.
  In this phase there is a high level of mixing between the $A$ and $B$ monomers.
  The reason is that, in the chosen setup of diblock copolymers~I, the $A$ and
  $B$ monomers are subjected to attractive forces.
  As a consequence, approximately
  in the range of temperatures $\bar T\le 0.7$,  the
  monomers belonging to different types
  form bonds in order to minimize 
  the energy of the knotted polymer. 
  The result is that the conformations are very compact and the monomers are
  closely packed together. This is visible
    in the colormap $A$  of Fig.~\ref{N200-phase-transition-colormaps}, left panel, where yellow is everywhere the dominant
  color.
  Of course, due to the excess of $A$ monomers,  which repel
  themselves, not all the
  $A$ monomers are able to bind with the limited number of $B$
  monomers available.
  As a result, small
  tails of $A$ monomers departing from the bulk are appearing, see Fig.~\ref{N200-phase-transition-sample-conformations},
  left panel. We associate these tails with the darker spots appearing in the colormap
  of Fig.~\ref{N200-phase-transition-colormaps}, left panel.
  From the colormap it turns out that the
  comparments $2-3$ and $17-20$
are responsible for the largest tail, as they
  are the most likely to be
  distant from each other. Indeed, the sector $x(2:6),y(17:20)$ is
  the darkest one.

At higher temperatures, in the interval between the two peaks of the
heat capacity of Fig.~\ref{N200-phase-transition} around $\bar
T_B=0.56$, knotted polymers with $N_A>>N_B$ are in the $\bar I-$phase \cite{NATFFMPLT2023}.
The colormap $B$ in Fig.~\ref{N200-phase-transition-colormaps}, central
panel, corresponds 
to the structural organization of the conformations typical of this phase.
The sector $x(31:40),y(1:30)$ is predominantly yellow-red as a
consequence of the fact that the $A$ monomers in compartments $1-30$
are forming bonds with the $B$ monomers in compartments $31-40$.
The result is a compact bulk held together by these bonds that is similar, but smaller than that of the
$\bar M-$phase. The level of mixing between the $A$ and
$B$ monomers is high.
The substantial difference from the previous phase is the large violet
area showing that the $A$ monomers in different compartments have a
lower chance to get close to each other. This is compatible with
the presence in the $\bar I-$phase of longer tails departing from
the compact bulk. These tails are the effect of the increased thermal
fluctuations that counteract the
attractive interactions between the $A$ and $B$ monomers. As a
consequence, the latter are no longer
able to sustain a large bulk as in the $\bar M-$phase and
consistent portions of the segment containing the $A$ monomers are
floating outside a smaller bulk. 
The polymer conformation in
Fig.~\ref{N200-phase-transition-sample-conformations}, central panel,
of energy $E=-110$
is indeed characterized by a compact bulk with long tails.

As the temperature increases further, the interactions between the monomers
are overwhelmed by the strong thermal fluctuations. As it is possible
to see in Fig.~\ref{N200-phase-transition-colormaps}, right panel, the
lighter yellow area is concentrated in the diagonal of colormap $C$,
implying that mainly the contiguous compartments are able to get close to each
other. The remaining area of the colormap is predominantly black or
violet. The fact that the violet color is concentrated in the sector
$x(31:40),y(1:30)$ has a simple explanation: It means that at the very high temperature $\bar T=5.00$ there is still some
residual effect of the attractive interactions between the $B$
monomers in compartments $31-40$ and the $A$ monomers in compartments $1-30$.
\subsection{Case of a circular $[4]$catenane}\label{4cat}
\begin{figure}[h]
  \begin{center}
\includegraphics[width=0.70\textwidth]{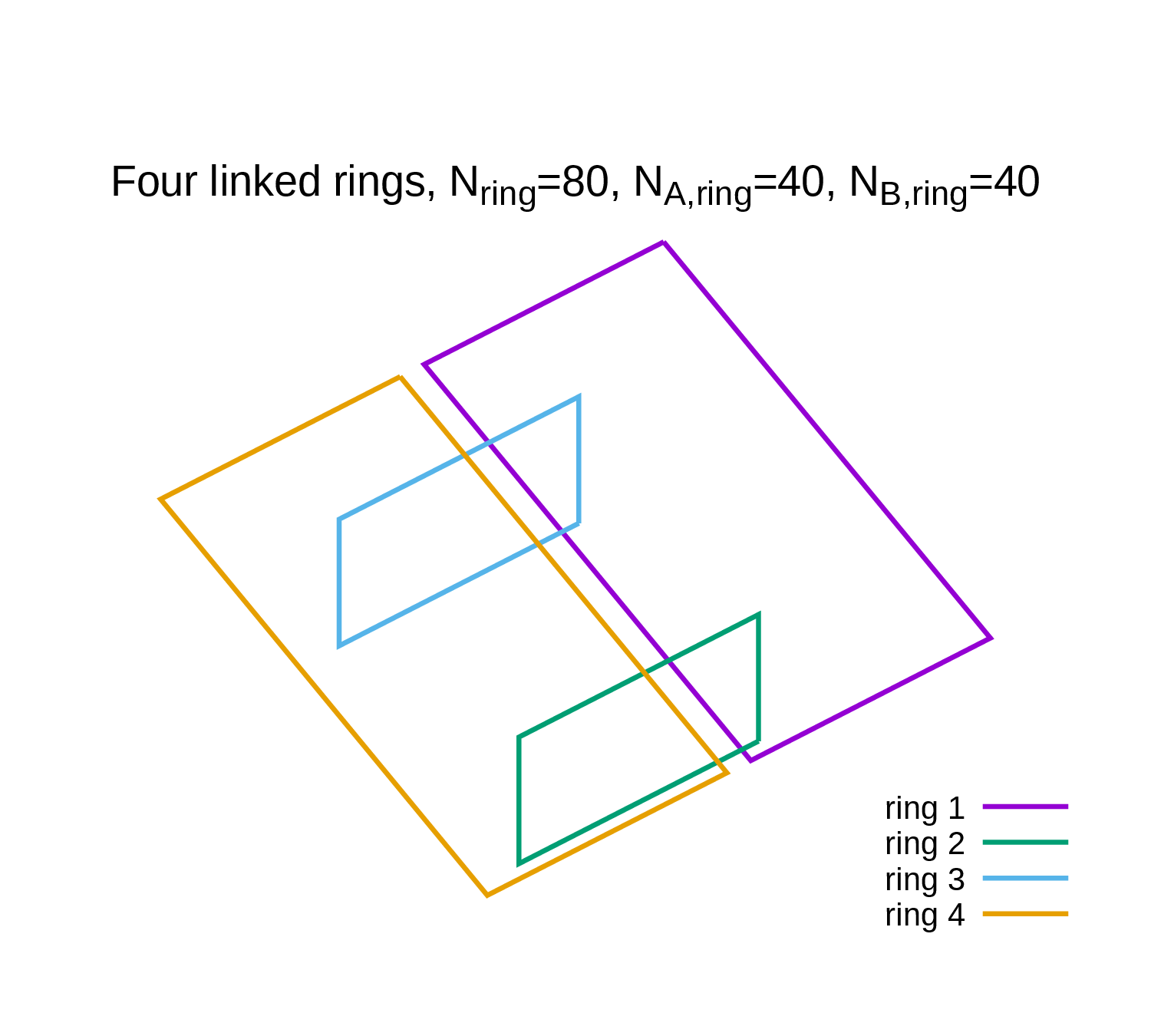}
  \end{center}
  \caption
      {This picture illustrates the topological structure of the circular $[4]$catenane discussed in this Subsection. Each ring composing the system contains $80$ monomers and is in the setup of diblock copolymer-II defined in Table~\ref{table1}. The diblock copolymer is composed by two segments containing $40$ monomers of type $A$ and $40$ monomers of type $B$. The solvent is good for the $A$ monomers and bad for the $B$ monomers.
  }
  \label{N320-seed}
\end{figure}

In this subsection, the contact matrices will be applied to study
the structural organizations at different temperatures
of the circular $[4]$catenane shown in Fig.~\ref{N320-seed}. This system is composed
by four polymer rings 
concatenated together of length $n_r=80$ each. The setup of each ring
is that of diblock copolymers~II, see Table~\ref{table1}, with
$n_{r,A}=40$ monomers of the $A$ type and $n_{r,B}=40$ monomers of the
$B$ type. The total number of
monomers in the system is $N=320$.
Each compartment contains $n_c=8$ monomers.
Accordingly, there are $40$ compartments in
the circular $[4]$catenane and the Hi-C matrix has dimension  $40\times 40=1600$.
Since each ring composing the link contains 10 compartments,
it is convenient to divide the Hi-C colormap into 16
sectors $s_{ab}$, $a,b=1,\ldots,4$, of dimension $10\times 10$.
For instance, $s_{34}$  denotes the sector x(21:30),y(31:40). 
The sectors $s_{ab}$ may be considered as smaller colormaps
providing information about the average locations of the compartments
of the $a-$th ring with respect to the compartments of the
$b-$th ring. The diagonal sectors
$s_{aa}$, $a=1,\ldots,4$ capture the structural organization
of the single rings.
It
will also be convenient to distinguish inside a
sector $s_{ab}$ the four
subsectors  $s_{ab}^{AA}$, $s_{ab}^{AB}$, $s_{ab}^{BA}$ and
$s_{ab}^{BB}$. For example, $s_{ab}^{BA}$ represents the colormap of
dimension $5\times 5$ which tells how
distant are in the average the five compartments of ring $a$ containing $B$
monomers from the five compartments containing
$A$ monomers of ring $b$.


\begin{figure}[h]
  \begin{center}
\includegraphics[width=0.95\textwidth]{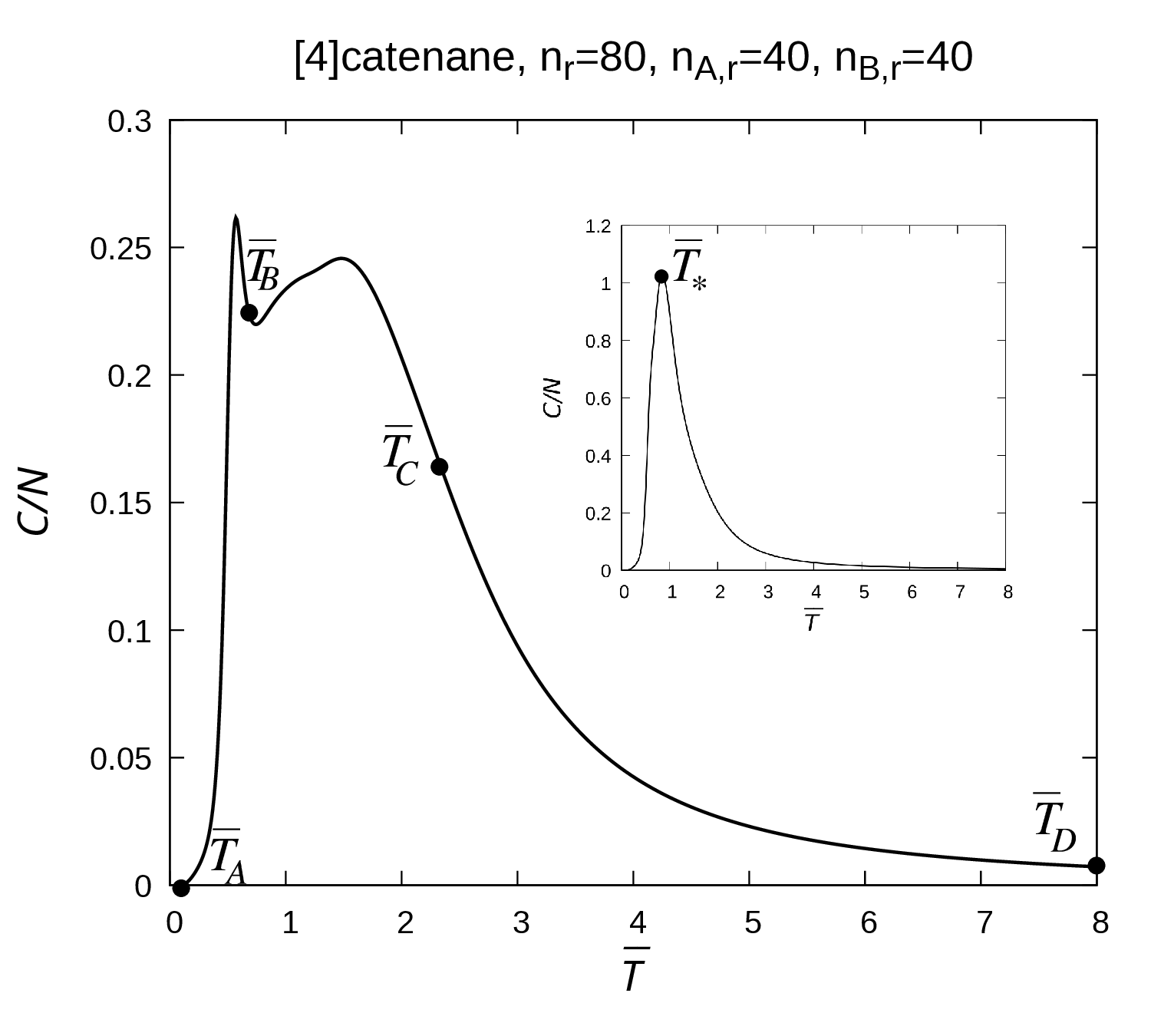}
  \end{center}
  \caption
{Shown in this picture is the plot of the specific heat capacity $C/N$ of the circular $[4]$catenane of Fig.~\ref{N320-seed} against the temperature $\bar T$. The first peak is centered at about $\bar T_{peak1}=0.56$. A shoulder is visible in the interval $[0.98,1.25]$. The last peak is at about $\bar T_{peak2}=1.25$. $\bar T_A,\ldots,\bar T_D$ are the temperatures at which the Hi-C colormaps of Figs.~\ref{N320-phase-transition-colormaps1} and \ref{N320-phase-transition-colormaps1} have been computed. In the inset the plot of the specific heat capacity of a single rings has been reported. It has a single peak centered at $\bar T_*\sim 0.85$.
  }
  \label{N320-phase-transition}
\end{figure}

The specific heat capacity of the circular $[4]$catenane is characterized by two
peaks and a shoulder at
$\bar T\sim 1.08$, see Fig.~\ref{N320-phase-transition}.
Following \cite{rampf,janke,janke2,hsieh,privalov}, peaks and shoulders appear in connections with phase transitions. This implies that the system admits four distinct phases and three phase transitions.
The most important features of the structural organization of the circular $[4]$catenane in all these four phases can be determined with the help of the Hi-C colormaps.
It turns our that there are two levels of organization: that of the single rings and that of the circular $[4]$catenane as a whole.

\begin{figure}[h]
  \begin{center}
\includegraphics[width=0.48\textwidth]{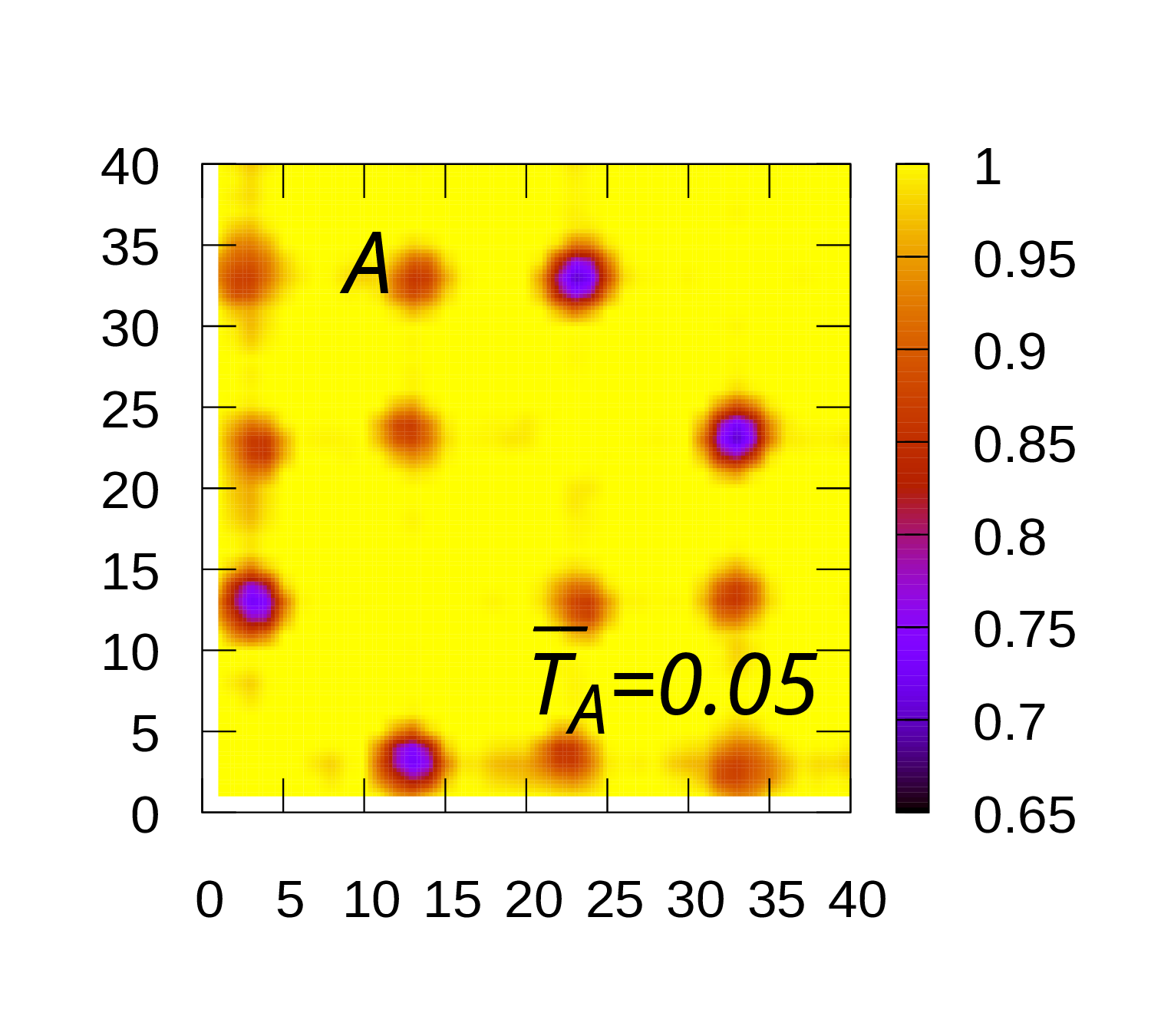}
\includegraphics[width=0.48\textwidth]{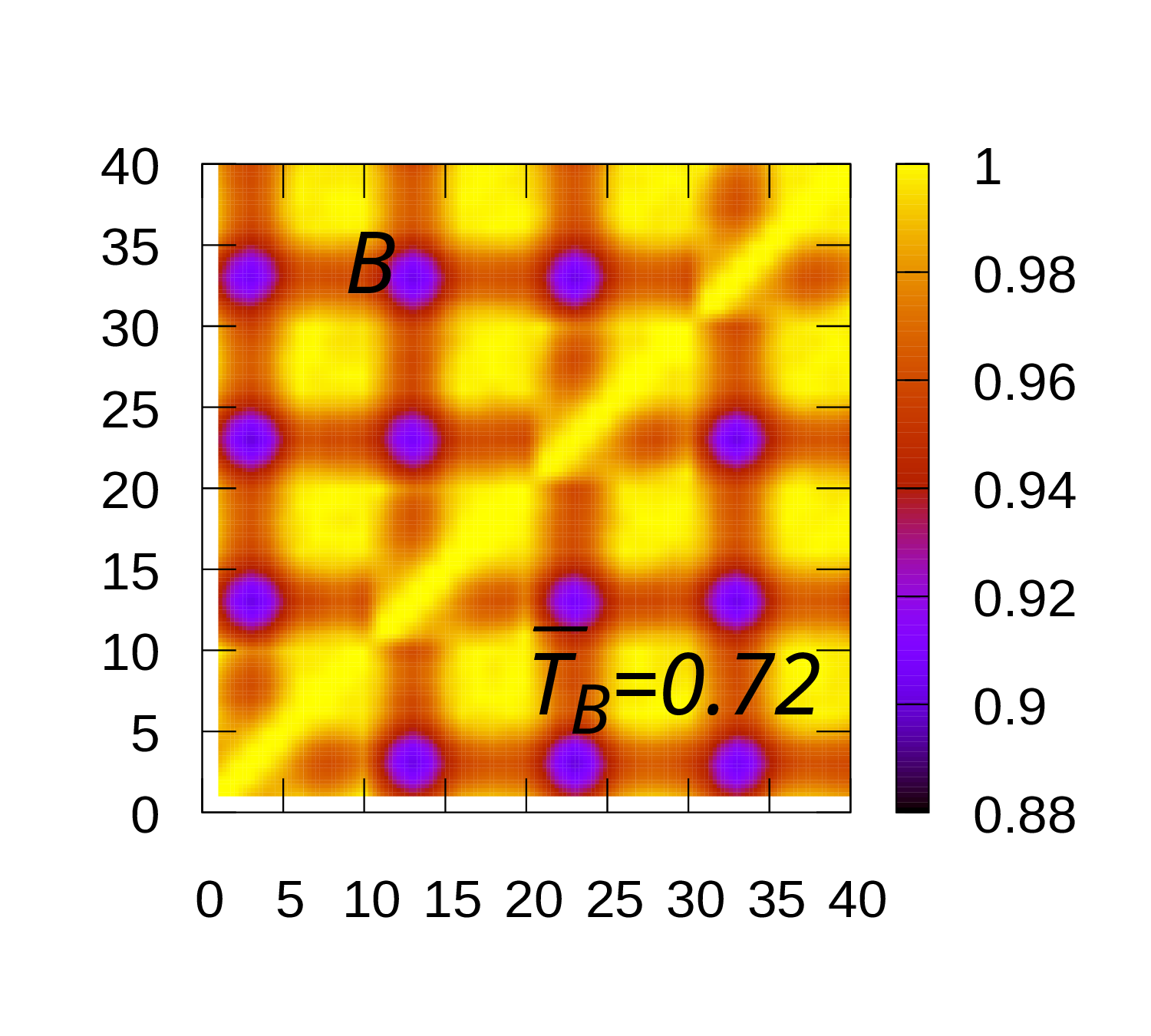}
  \end{center}
  \caption
{This figure shows the colormaps of the circular $[4]$catenane at the temperatures $\bar T_A=0.05$ (left panel) and $\bar T_B=0.72$ (right panel).
  }
  \label{N320-phase-transition-colormaps1}
\end{figure}

\begin{figure}
  \begin{center}
    \includegraphics[width=0.48\textwidth]{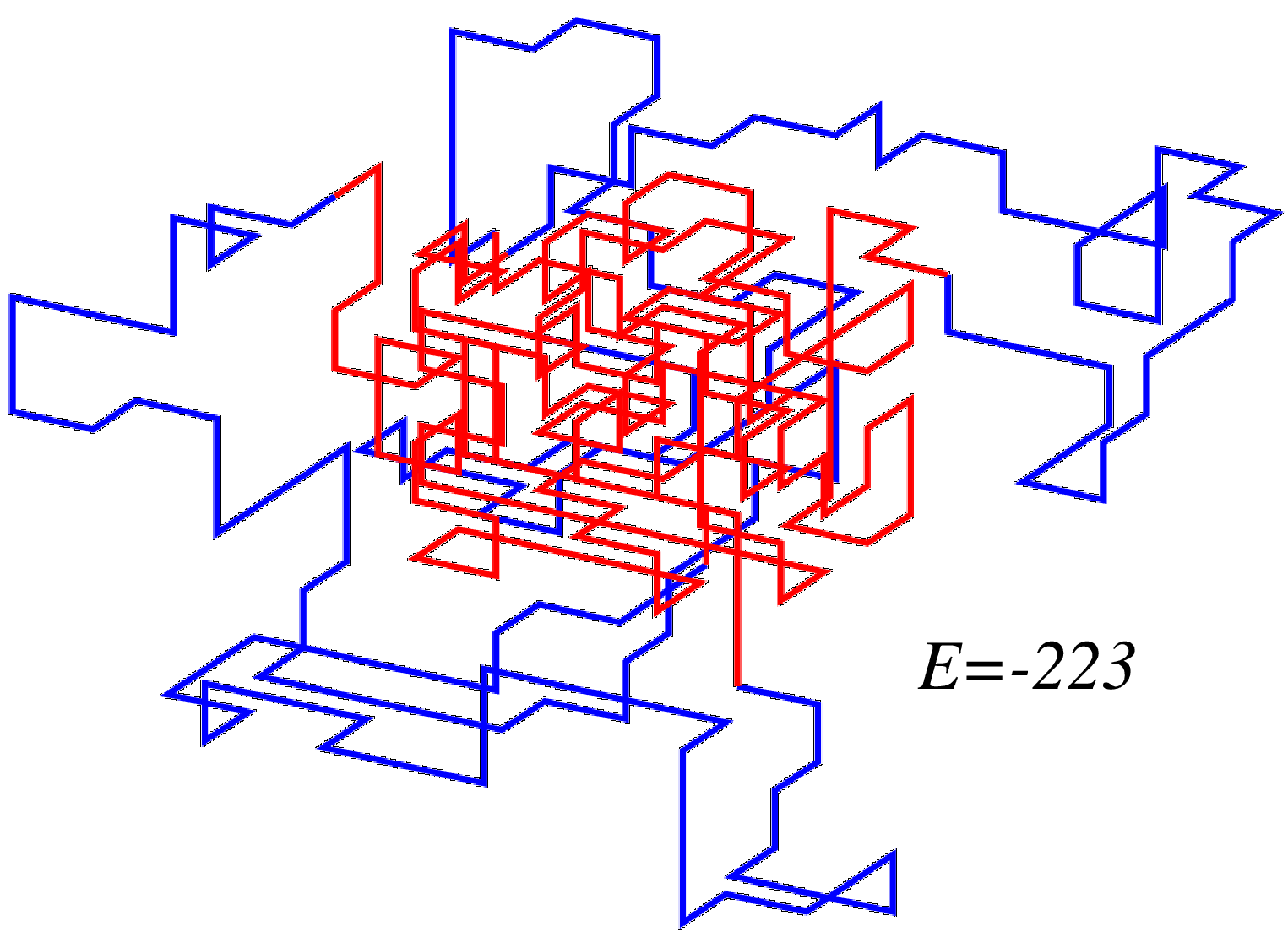}
    \includegraphics[width=0.48\textwidth]{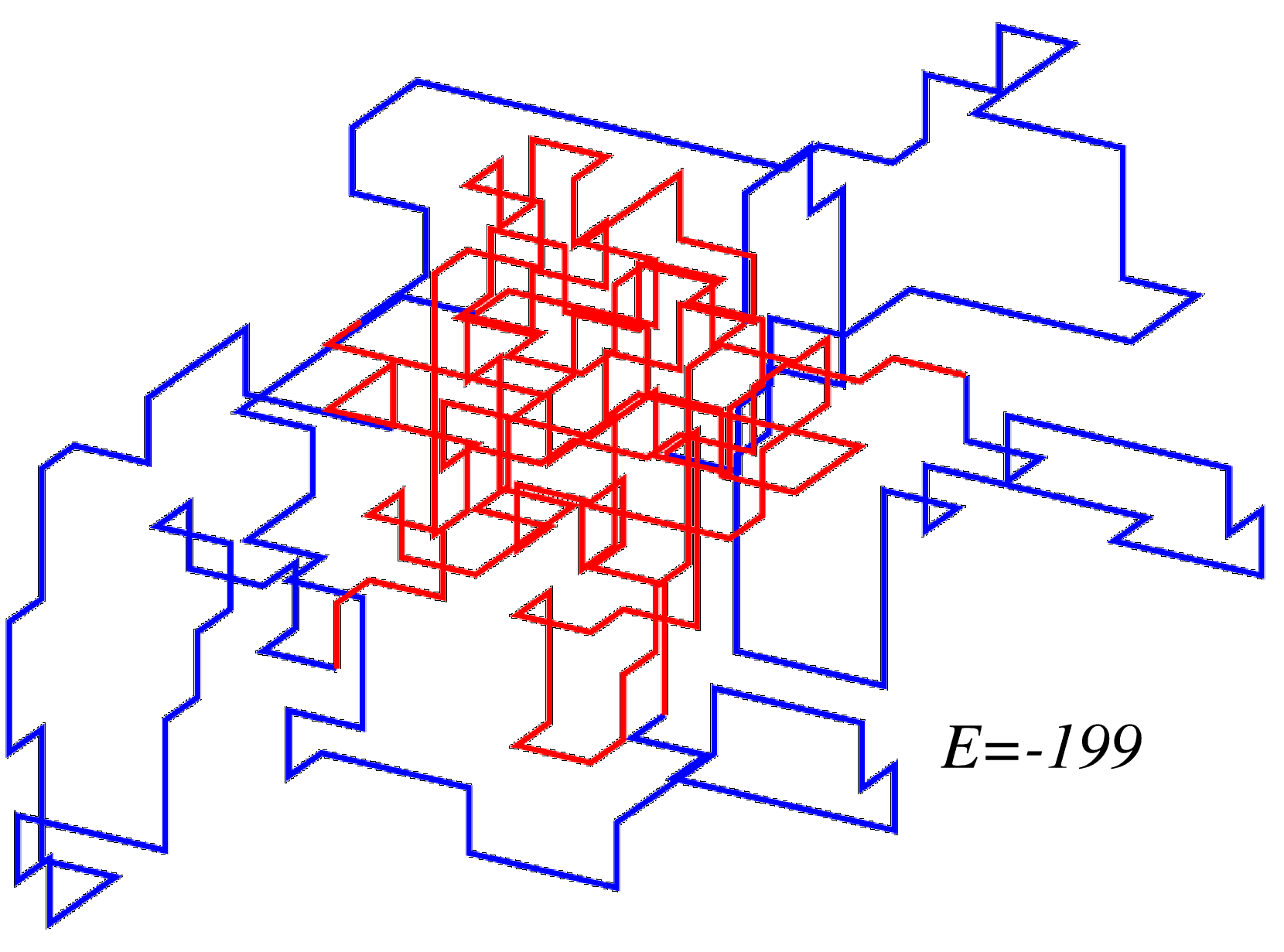}
  \end{center}\caption{
    This Figure shows snapshots of conformations with energies $E=-223$ (left panel) and $E=-199$. $E=-223$ and $E=-199$ are approximately the average energies of the circular $[4]$catenane at the temperatures $\bar T_A=0.05$ and  $\bar T_B=0.72$ respectively. The $A$ monomers are located in the blue segments and the $B$ monomers in the red segments. For convenience, the rings have not been distinguished.
  }\label{N320-snapshots1}
 \end{figure}

The first phase $p_1$ appears at the lowest temperatures.
In the colormap taken at $\bar T_A=0.05$ in
Fig.~\ref{N320-phase-transition-colormaps1},
left panel, yellow is the dominant color. Since yellow corresponds in the colorbar to the highest probability
that two compartments are found close to each other, it is possible to conclude that  the rings composing the circular $[4]$catenane  as well as the whole system are in compact conformations.
As already mentioned, the information about the $a-$th ring is stored in the sector $s_{aa}$, $a=1,\ldots,4$ which is  mainly colored with bright yellow. This means that the rings are in a collapsed phase. This could be expected in the case of the compartments containing the $B$ monomers because in the chosen setup diblock copolymer~II they
are subjected to attractive interactions. More surprisingly is that also the compartments with the $A$ monomers, which repel themselves, can be found near to each other as it is easy to realize by looking at the subsectors $s_{aa}^{AA}$, $i=1,\ldots,4$ of the colormap
of Fig.~\ref{N320-phase-transition-colormaps1}, left panel. All these subsectors are indeed colored with yellow. This is however not a contradiction, because only the formation of bonds can change the energy according to the Hamiltonian (\ref{hamI}). The condition (\ref{contactcondition}) for two compartments to become close does not necessarily imply that the monomers of these compartments are forming bonds. Concerning the global structure of the circular $[4]$catenane, we notice that yellow is the dominant color in
the sectors outside the diagonal, i.~e. $s_{ab}$ with $a\ne b$.
The fact that the subsectors $s_{ab}^{BB}$ with $a\ne b$ are yellow suggest that the compartments with the $B$ monomers are densely packed together due to the attractive interactions and form a compact bulk within the circular $[4]$catenane.
Instead, the subsectors $s_{ab}^{AA}$ with $a\ne b$ 
are much darker than all the other subsectors.
To fix the ideas, let's consider
in the colormap of Fig.~\ref{N320-phase-transition-colormaps1}, left panel,
the three dark spots that are visible in the sector x(1:5),y(1:40). The presence of the first spot in the subsector $s_{12}^{AA}$ in purple color reveals that, in the average,
the lowest probability to find two non-contiguous compartments close to each other occurs in the case of the compartments with the $A$ monomers of the first ring and the compartments with the $A$ monomers of the second ring. The  brighter red spots in subsectors $s_{13}^{AA}$ and $s_{14}^{AA}$ are telling instead that this probability is higher for the compartments of the $A$ monomers of the first ring and those of the third and fourth rings.
Here the first ring has been singled out, but of course the situation is symmetric if select the second ring  starting from the sector x(11:15),y(1:40)) or the remaining rings $3$ and $4$, see the sectors x(21:25),y(1:40)) and x(31:35),y(1:40) respectively. In summary, the colormap of Fig.~\ref{N320-phase-transition-colormaps1}, left panel, obtained by considering hundred of billions of conformations, suggests that the structure of an overwhelming number of these conformations at the lowest temperatures is of the kind shown in Fig.~\ref{N320-snapshots1}, left panel: The $B$ monomers (in red) of all the rings are tightly packed together while the $A$ monomers (in blue), that are subjected to repulsive interactions, are organized in four long tails that stay far from each other in order to minimize the energy of the circular $[4]$catenane. The signature of these four tails are the twelve darker spots visible in the colormap of Fig.~\ref{N320-phase-transition-colormaps1}, left panel.
An example of conformations in the $p_1$ phase has been given in  Fig.~\ref{N320-snapshots1}, left panel.

With raising temperatures, the circular $[4]$catenane has a first phase transition $p_1\longrightarrow p_2$ at about $\bar T\sim 0.58$, i. e. the temperature of the first peak of the specific heat capacity in Fig.~\ref{N320-phase-transition}.
In the phase $p_2$ the conformations are very similar to those in $p_1$, see the colormap in Fig.~\ref{N320-phase-transition-colormaps1}, right panel computed at $\bar T_B=0.72$. There are namely twelve darker spots corresponding to the four tails mentioned before and the four rings are still held strongly together by the bonds formed by the $B$ monomers.
However, the conformations of the single rings as well as those of the whole system are not tight as in the phase $p_1$.  This can be seen by looking for instance at the subsectors $s_{aa}^{AA}$ which are yellow mainly near the diagonal line, while the upper and lower triangles are somewhat darker.
Moreover, the compartments with the $A$ monomers are no longer sticking near the compartments with the $B$ monomers as it is in the phase $p_1$. Indeed,  all subsectors $s_{ab}^{AB}$ and $s_{ab}^{BA}$ in Fig.~\ref{N320-phase-transition-colormaps1}, right panel, that are
yellow in the colormap of the left panel corresponding to the $p_1$ phase, are now reddish.
The fact that the conformations in $p_2$ are more loosely packed than in $p_1$ implies that the monomers have an increased mobility. Indeed, the range
between $0.65$ and $1$
in the colorbar of Fig.~\ref{N320-phase-transition-colormaps1}, left panel, is bigger that the range in the right panel which goes from $0.88$ to $1$. As previously mentioned, the range of values in the colorbars is dependent on the presence of constraints or on the number of conformations that are accessible in a given phase.
The typical conformation in the $p_2$ phase is shown in  Fig.~\ref{N320-snapshots1}, right panel.

\begin{figure}[h]
  \begin{center}
\includegraphics[width=0.48\textwidth]{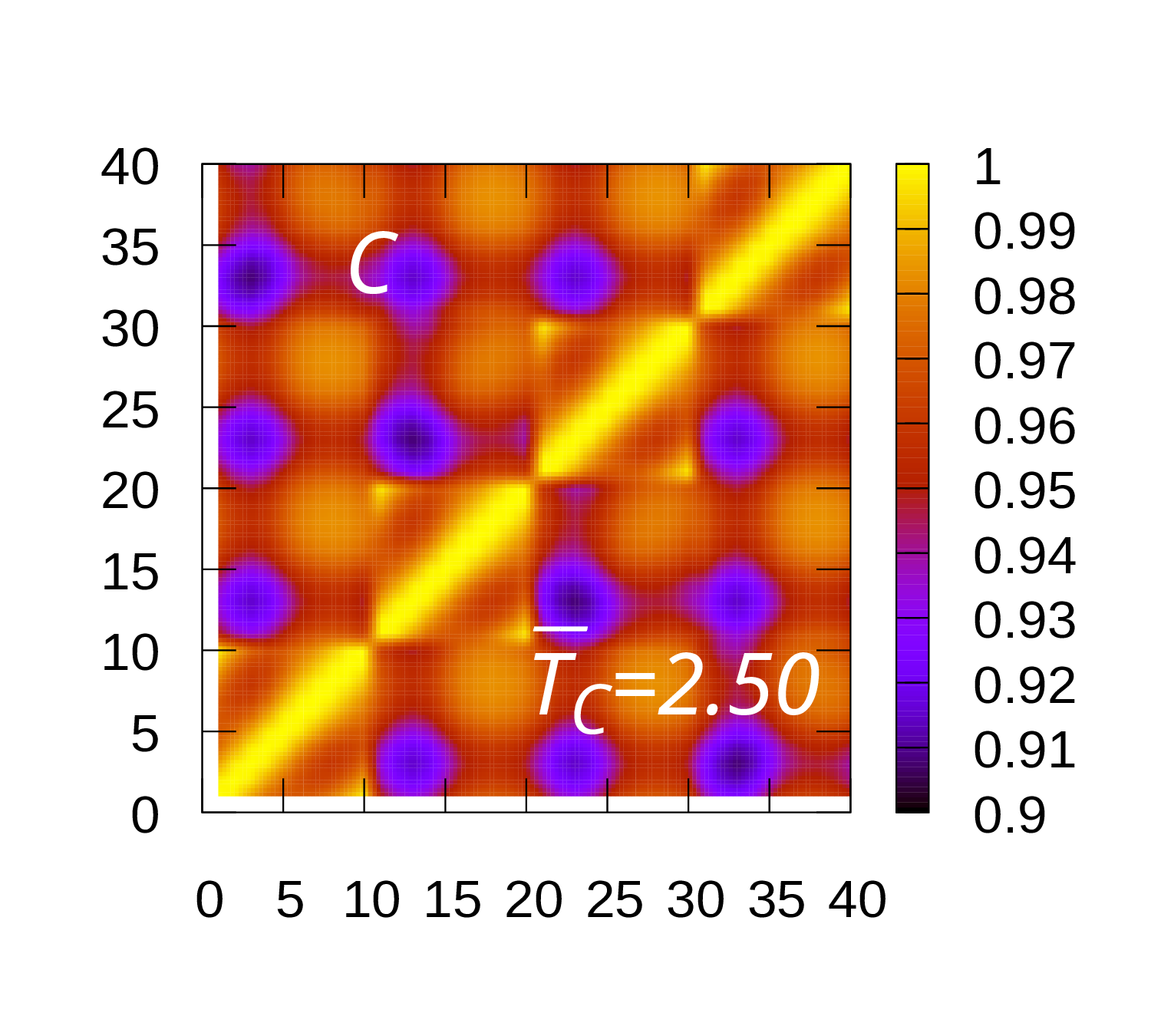}
\includegraphics[width=0.48\textwidth]{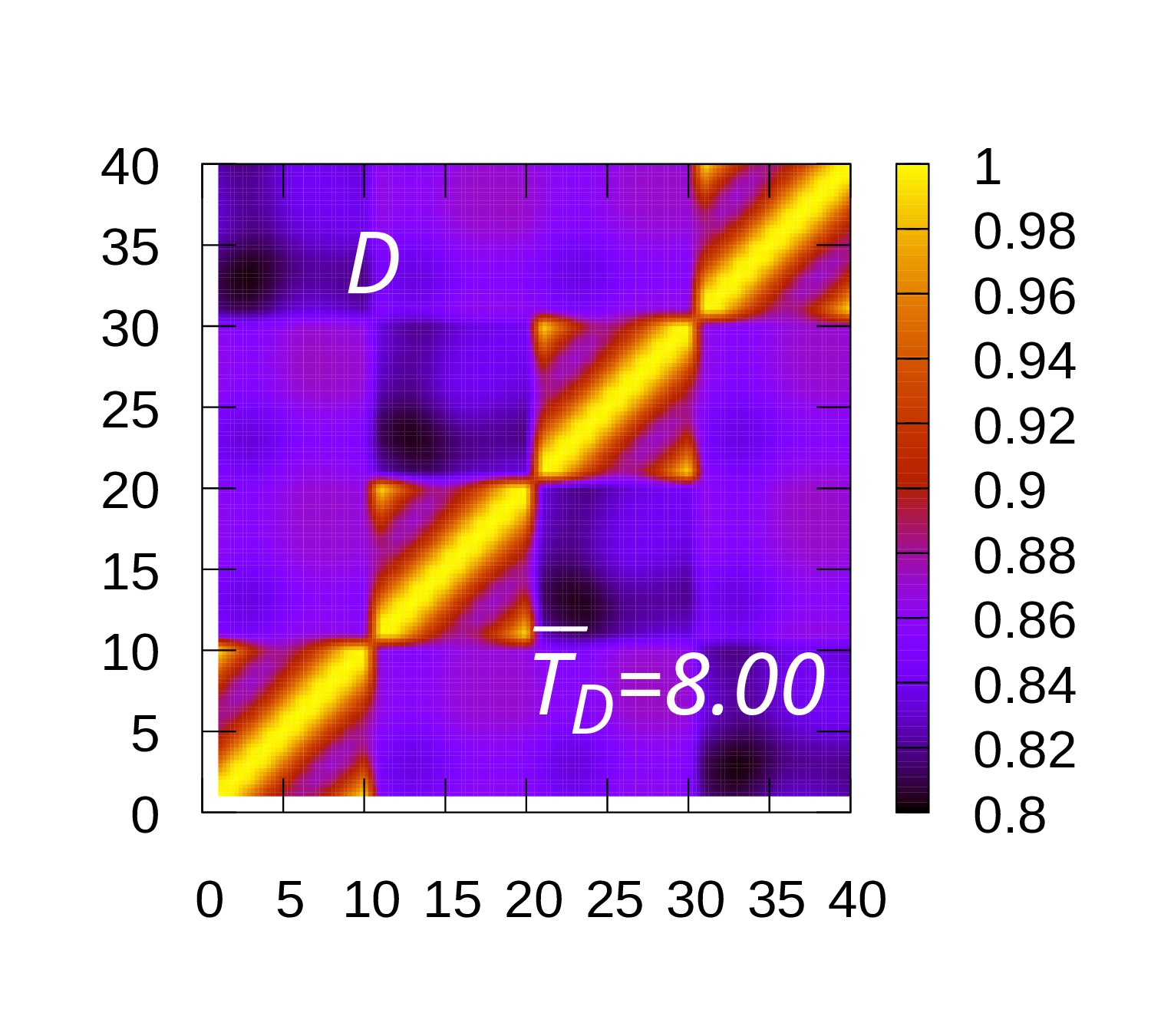}
  \end{center}
  \caption
{This figure shows the colormaps of the circular $[4]$catenane at the temperatures $\bar T_C=2.50$ (left panel) and $\bar T_D=8.00$ (right panel).
  }
  \label{N320-phase-transition-colormaps2}
\end{figure}

The next phase is $p_3$. The main properties of its conformations are encoded in the colormap of Fig.~\ref{N320-phase-transition-colormaps2}, left panel.
The temperature of the colormap is $\bar T_C=2.50$
Also this phase is characterized by four tails, whose signature are the twelve darker spots as explained before. The yellow color is dominant only in along the diagonal line of the colormap, showing that the contiguous compartments are the most likely to be found close to each other. The sectors $s_{ab}^{BB}$ with $a,b=1,\ldots,4$ and $a\ne b$ are still colored with yellow like in the $p_1$ and $p_2$ phases. Even if the tone of this yellow is darker, this means that also at the relatively high temperature $\bar T_C=2.50$ the circular $[4]$catenane is in a compact conformation in which all the four rings are held together by the attractive interactions between the $B$ monomers.
A difference from the previously discussed phases is that the single rings are now in an expanded and unmixed state~\footnote{Let's notice that a single rings alone makes the transition from the compact/mixed phase to the expanded/unmixed phase at much lower temperatures, with the peak of the transition occurring at $\bar T_*\sim 0.85$, see Fig.~\ref{N320-phase-transition}.}. The diagonal sectors $s_{aa}$ are in fact yellow colored only in the neighborhood of the diagonal line while the upper and lower triangles are darker. This is the typical signature of an unmixed/expanded ring in which the interactions between the monomers
become negligible in comparison with the strong thermal fluctuations. A similar behavior has already been found in the $\bar U-$phase of the $4_1$ knot, see Fig.~\ref{N200-phase-transition-colormaps}, right panel and related comments.


\begin{figure}
  \begin{center}
    \includegraphics[width=0.48\textwidth]{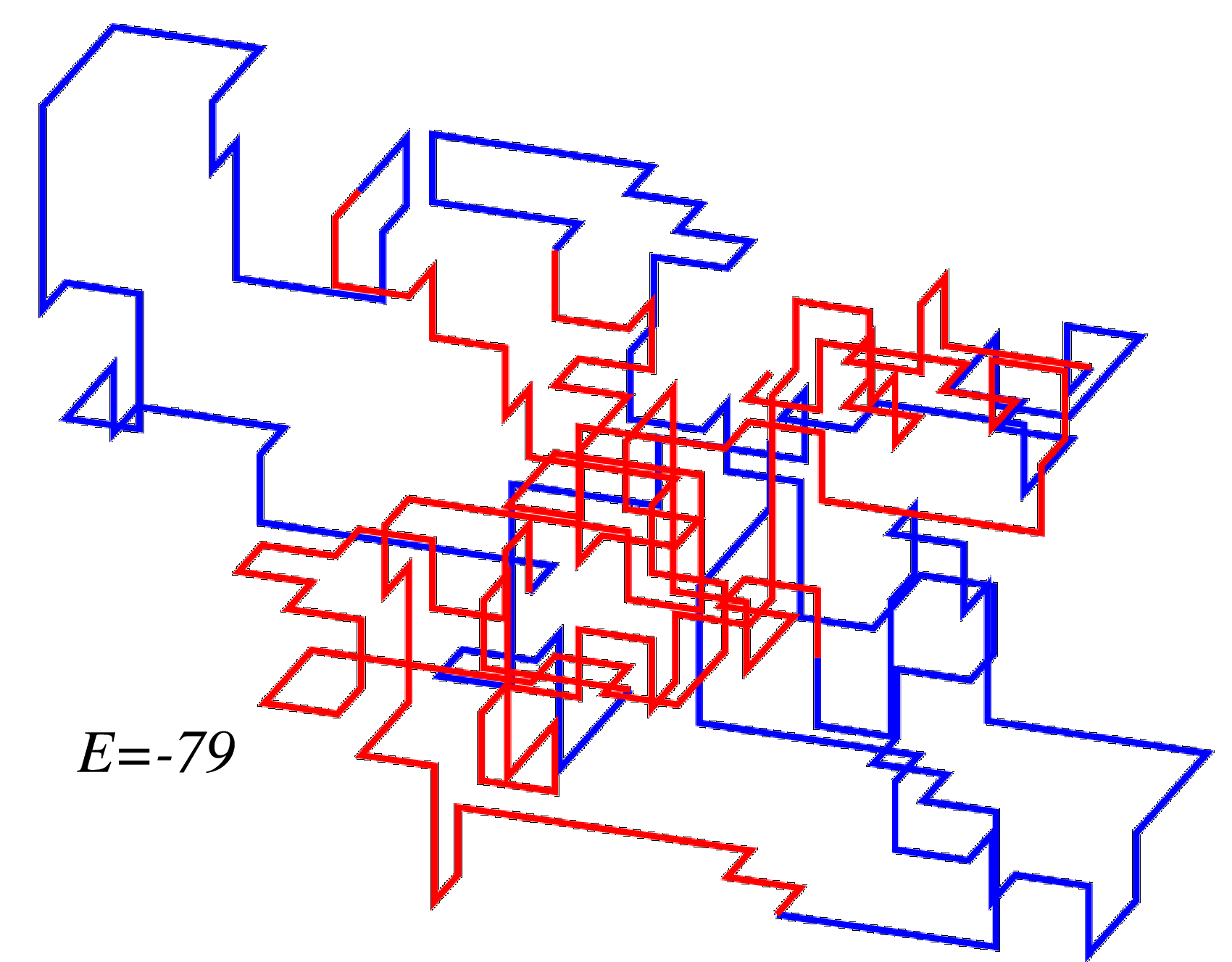}
    \includegraphics[width=0.48\textwidth]{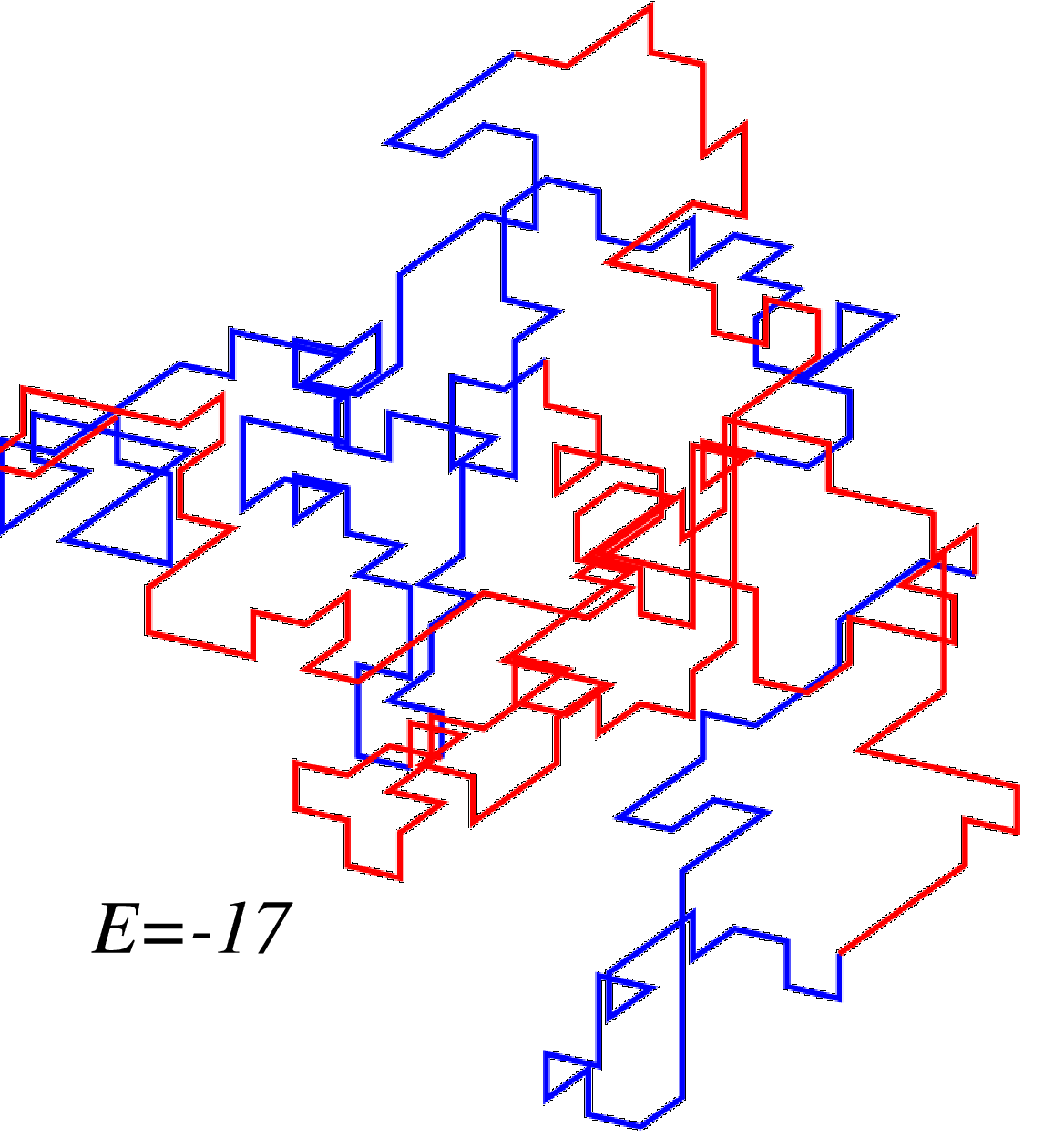}
  \end{center}\caption{
This Figure shows snapshots of conformations with energies $E=-79$ (left panel) and $E=-17$. $E=-79$ and $E=-17$ are approximately the average energies of the circular $[4]$catenane at the temperatures $\bar T_C=2.50$ and  $\bar T_D=8.00$ respectively. The $A$ monomers are located in the blue segments and the $B$ monomers in the red segments. For convenience, the rings have not been distinguished.
  }\label{N320-snapshots2}
 \end{figure}

Finally, in the last phase $p_4$, which is dominant at the highest temperatures, the interactions become negligible due to the strong thermal fluctuations. In this situation, only the contiguous compartments have the highest chance to get close. This is why the colormap of Fig.~\ref{N320-phase-transition-colormaps2}, right panel, has dark tones apart from the diagonal.
These tones are darker in correspondence of the sectors $s_{ab}^{AA}$ and lighter in the case of the sectors $s_{ab}^{BB}$. This can be explained by the fact that the attractive interactions between the $B$ monomers and the repulsive ones between the $A$ monomers are still playing a role.
The stuctural organization into a bulk held together by the $B$ monomers and the four tails has however been lost. A conformation that well summarizes these characteristics is shown in Fig.~\ref{N320-snapshots2}, right panel.

\section{Conclusions}\label{conclusions}
The plots of the heat capacity of knotted polymers and polycatenanes
show that these systems undergo several phase transition, but it is often not easy to identify the differences in the structural organization of the conformations that characterize the various phases.
The Hi-C inspired method introduced in this work is very useful in this respect, because the Hi-C matrices and the related colormaps
are able to capture the relevant features that distinguish a given phase from the others.

The method has been illustrated here in the particular cases of a knotted diblock copolymer with the topology of the knot $4_1$ and a circular polycatenane consisting of four rings linked together, but several other topologies have been tested.
There is an outstanding agreement between the number of phases predicted by the plots of the specific heat capacity and the number of patterns shown by the colormaps. 
The results concerning the knotted diblock copolymer confirm the previous findings of Ref.~\cite{NATFFMPLT2023}, but also provide further information.
For instance, they prove that the tails observed in the conformations analysed in \cite{NATFFMPLT2023} are a characteristics of the mixed $\bar M$ and intermediate $\bar I$ phases. In particular,
the signature provided by the darker spots in the colormap
of Fig.~\ref{N200-phase-transition-colormaps}, left panel, suggests
that the conformations of the knotted copolymer like that of Fig.~\ref{N200-phase-transition-sample-conformations}, left panel, with three small tails, one of which somewhat bigger than the other two, are common
at the lowest temperatures.
We have also seen that the range of the values in the colorbars is related to the mobility of the monomers and can be bigger or smaller depending on the presence of constraints or not.
The analysis of the circular $[4]$catenane has revealed that the phase transitions are related to rearrangements of the polymer at two scales: that of the single rings composing the polycatenane and that of the whole system.
The sectors along the diagonal of the main colormap are colormaps themselves providing information about the conformations of the ring. The sectors outside the diagonal describe how frequently the compartments belonging to different rings may be found in close vicinity in randomly generated conformations of the circular $[4]$catenane.

The Hi-C inspired method relies on the condition of Eq.~(\ref{contactcondition}) that is simple to implement in a simulation without significantly increasing the time necessary for the calculations.
The resolution of the polymer structure is determined by the parameter $n_c$ that fixes the number of monomers in a compartment. The results are
robust with respect to the change of this resolution, see Fig.~\ref{colormap} and related comments.
The set of Hamiltonians (\ref{hamI}) used to model the systems discussed in this paper implies short interactions.
Work is in progress to extend the results to other types of interactions.
\begin{acknowledgments} 
The simulations reported in this work were performed in part using the HPC
cluster HAL9000 of
the University of Szczecin.
The research presented here has been supported by the Polish National Science Centre under
grant no. 2020/37/B/ST3/01471.
This work results within the collaboration of the COST
Action CA17139 (EUTOPIA).
L.T.
acknowledges financial support from ICSC-Centro Nazionale
di Ricerca in High Performance Computing, Big Data
and Quantum Computing, funded by European Union-
NextGenerationEU. 
The use of some of the facilities of the Laboratory of
Polymer Physics of the University of Szczecin, financed by 
a grant of the European Regional Development Fund in the frame of the
project eLBRUS (contract no. WND-RPZP.01.02.02-32-002/10), is
gratefully acknowledged.  \end{acknowledgments}

\end{document}